\documentclass[amsmath,amssymb,aps,twocolumn,superscriptaddress,longbibliography,notitlepage]{revtex4-1}
\usepackage{graphicx}
\usepackage{verbatim}
\usepackage{amsmath}
\usepackage{subfigure}
\usepackage{hyperref}

\begin{document}

\title{Experimental and Theoretical Investigation of the Crossover from the Ultracold to the Quasiclassical Regime of Photodissociation}

\author{I. Majewska}
\affiliation{Quantum Chemistry Laboratory, Department of Chemistry, University of Warsaw, Pasteura 1, 02-093 Warsaw, Poland}
\author{S. S. Kondov}
\affiliation{Department of Physics, Columbia University, 538 West 120th Street, New York, NY 10027-5255, USA}
\author{C.-H. Lee}
\affiliation{Department of Physics, Columbia University, 538 West 120th Street, New York, NY 10027-5255, USA}
\author{M. McDonald}
\altaffiliation{Present address:  Department of Physics, University of Chicago, 929 East 57th Street GCIS ESB11, Chicago, IL 60637, USA}
\affiliation{Department of Physics, Columbia University, 538 West 120th Street, New York, NY 10027-5255, USA}
\author{B. H. McGuyer}
\altaffiliation{Present address:  Facebook, Inc., 1 Hacker Way, Menlo Park, CA 94025, USA}
\affiliation{Department of Physics, Columbia University, 538 West 120th Street, New York, NY 10027-5255, USA}
\author{R. Moszynski}
\affiliation{Quantum Chemistry Laboratory, Department of Chemistry, University of Warsaw, Pasteura 1, 02-093 Warsaw, Poland}
\author{T. Zelevinsky}
\email{tanya.zelevinsky@columbia.edu}
\affiliation{Department of Physics, Columbia University, 538 West 120th Street, New York, NY 10027-5255, USA}

\begin{abstract}
At ultralow energies, atoms and molecules undergo collisions and reactions that are best described in terms of quantum mechanical wave functions.  In contrast, at higher energies these processes can be understood quasiclassically.  Here, we investigate the crossover from the quantum mechanical to the quasiclassical regime both experimentally and theoretically for photodissociation of ultracold diatomic strontium molecules.  This basic reaction is carried out with a full control of quantum states for the molecules and their photofragments.  The photofragment angular distributions are imaged, and calculated using a quantum mechanical model as well as the WKB and a semiclassical approximation that are explicitly compared across a range of photofragment energies.  The reaction process is shown to converge to its high-energy (axial recoil) limit when the energy exceeds the height of any reaction barriers.  This phenomenon is quantitatively investigated for two-channel photodissociation using intuitive parameters for the channel amplitude and phase.  While the axial recoil limit is generally found to be well described by a commonly used quasiclassical model, we find that when the photofragments are identical particles, their bosonic or fermionic quantum statistics can cause this model to fail, requiring a quantum mechanical treatment even at high energies.
\end{abstract}

\maketitle

\newcommand{\Schematic}{
\begin{figure}
\includegraphics*[trim = 1.7in 3in 1.7in 0.5in, clip, width=3.375in]{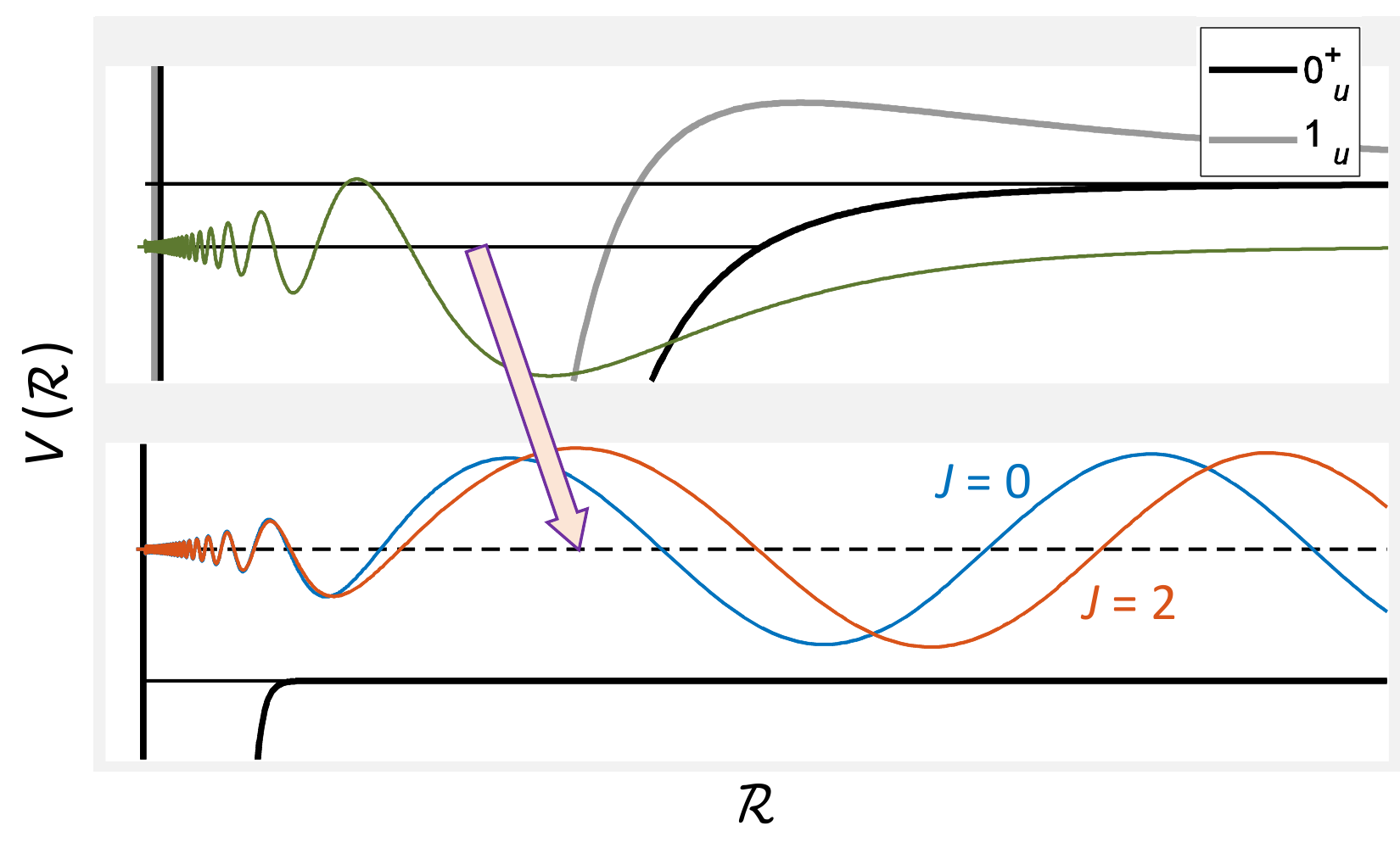}
\caption{Molecular potentials for the $^{88}$Sr$_2$ electronic states used in this work.  The ground state $X0_g^+$ correlates to the $^1S+{^1S}$ atomic threshold while the excited states $0_u^+$ and $1_u$ correlate to $^1S+{^3P_1}$.  The $1_u$ potential has a $\sim1$ mK electronic barrier.  The molecules are photodissociated via an electric-dipole optical transition from bound states of mostly $0_u^+$ or $1_u$ character to the ground-state continuum.  A range of rovibrational initial states are explored.  In this example, weakly bound $0_u^+(v=-2,J_i=1,M_i=0)$ molecules are photodissociated, and the photofragments occupy two allowed partial waves $J=\{0,2\}$ shown at the energy of 50 MHz (2.4 mK).}
\label{fig:Schematic}
\end{figure}
}

\newcommand{\PADs}{
\begin{figure*}
\includegraphics*[trim = 0.2in 3.2in 0.2in 0.2in, clip, width=7in]{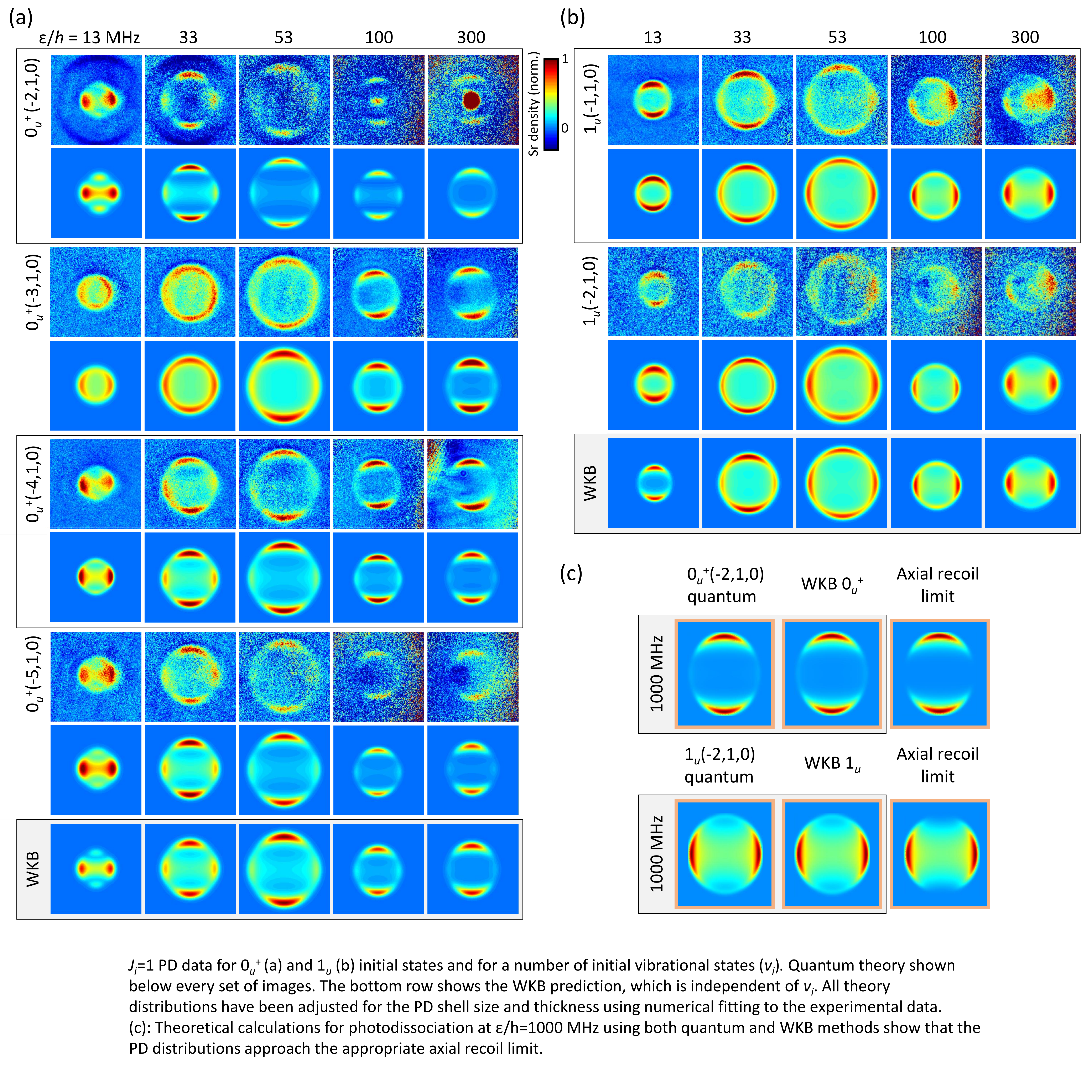}
\caption{Measured and calculated photofragment angular distributions in the ultracold quantum mechanical regime and in the high-energy axial recoil limit.  (a) The $0_u^+$ initial states are explored at the continuum energies $\varepsilon/h=\{13,33,53,100,300\}$ MHz ($\varepsilon/k_B=$0.6-14 mK).  For each initial state, the top and bottom rows correspond to measurements and quantum mechanical theory, respectively.  The \textit{ab initio} images were adjusted for the shell size and thickness in experimental data, where the photofragment ring diameter is typically 0.3-0.5 mm.  The strong dependence of the patterns on $\varepsilon$ and the less pronounced dependence on $v$ are discussed in the text.  The lowest row shows $v$-independent distributions obtained with the WKB approximation.  The most weakly bound state exhibits central dots that arise from spontaneous photodissociation into the optical lattice; these are more apparent at higher energies where the signal is weaker, and can be ignored.  (b) Same as (a), for $1_u$ initial molecular states.  (c) Angular distributions calculated for a pair of $0_u^+$ and $1_u$ weakly bound states using both quantum theory and the WKB approximation at a high energy $\varepsilon/h=1000$ MHz ($\varepsilon/k_B=48$ mK), to show their close agreement with the appropriate axial recoil limit.
}
\label{fig:PADs}
\end{figure*}
}

\newcommand{\RParametersV}{
\begin{figure}
\includegraphics*[trim = 5.5in 0.5in 2in 0.5in, clip, width=3.375in]{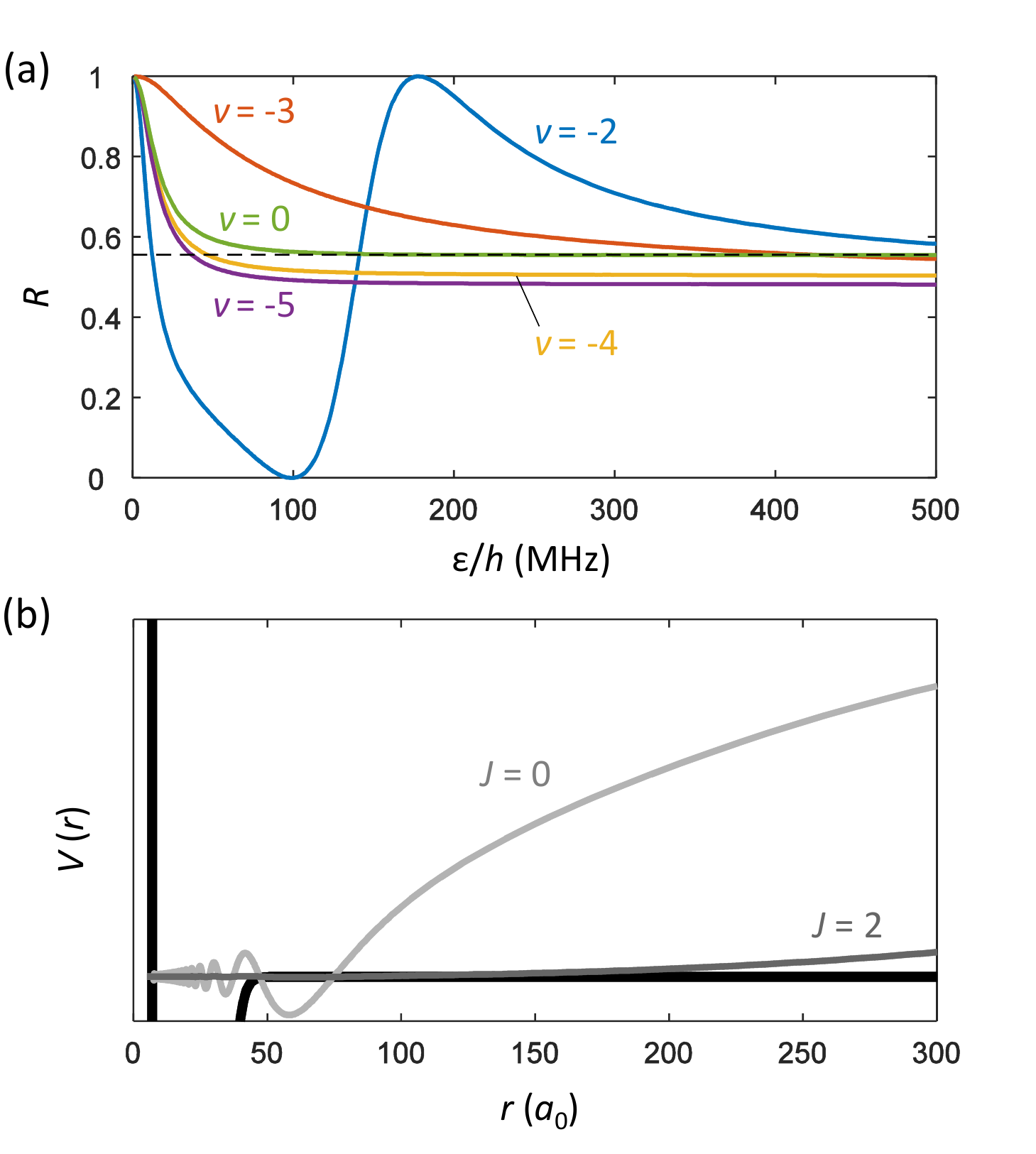}
\caption{(a) Quantum mechanical calculation of the amplitude-squared $R$ parameter for two-channel photodissociation as in Eq. (\ref{eq:IQM2J}).  The initial molecular state is $0_u^+(v,1,0)$ with $v=\{-2,-3,-4,-5\}$, and $P=0$.  For comparison, the most deeply bound $v=0$ state is also shown.  The axial recoil limit is indicated with a dashed line.  (b) The allowed continuum wave functions with $J=\{0,2\}$ are plotted for very low energy $\varepsilon/h=0.5$ MHz ($\varepsilon/k_B=24$ $\mu$K).  The negligible contribution of the $J=2$ partial wave due to its repulsive rotational barrier explains why $R=0$ as $\varepsilon\rightarrow0$ in (a) and thus only the lowest allowed partial wave is populated in near-threshold photodissociation.}
\label{fig:RParametersV}
\end{figure}
}

\newcommand{\RParameterBarrierHeight}{
\begin{figure}
\includegraphics*[trim = 3in 4in 2.5in 0.5in, clip, width=3.375in]{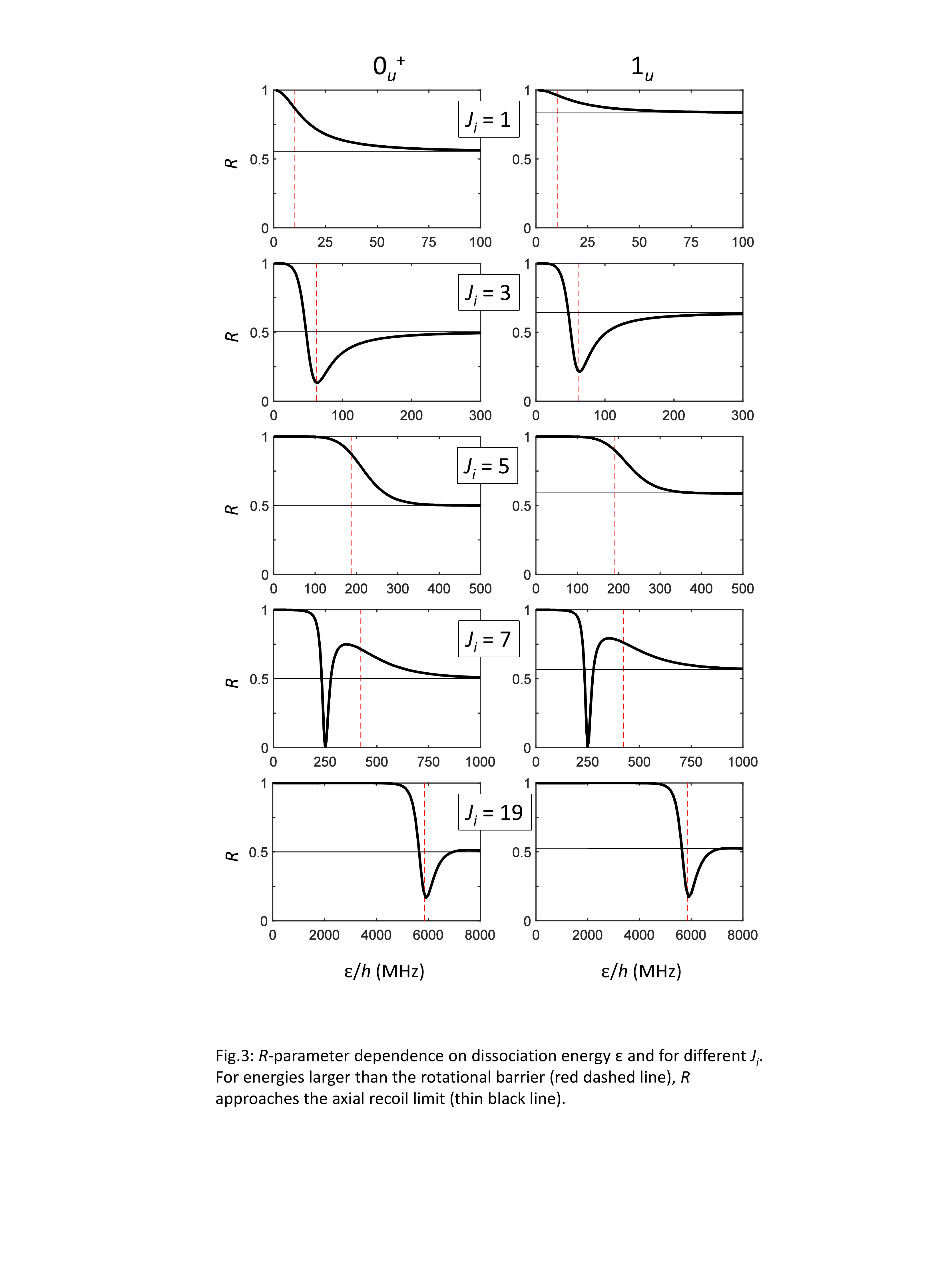}
\caption{Dependence of the amplitude-squared $R$ parameter from Eq. (\ref{eq:IQM2J}) on energy and on initial angular momentum $J_i$.  The initial molecular states are $0_u^+(0,J_i,0)$ (left column) and $1_u(0,J_i,0)$ (right column), and $P=0$.  In each frame, the axial recoil limit is indicated by a horizontal line, and the highest rotational barrier in the continuum is shown by a dashed vertical line.  For larger $J_i$, the axial recoil limit is approached at higher energies, but in all cases the crossover from the ultracold to the quasiclassical regime occurs at the energy scale set by the barrier height.  The initial states with $J_i=3,7,19$ exhibit shape resonance behavior.}
\label{fig:RParameterBarrierHeight}
\end{figure}
}

\newcommand{\QuantumWKBSemiDelta}{
\begin{figure}
\includegraphics*[trim = 1.6in 3in 1.6in 0.4in, clip, width=3.375in]{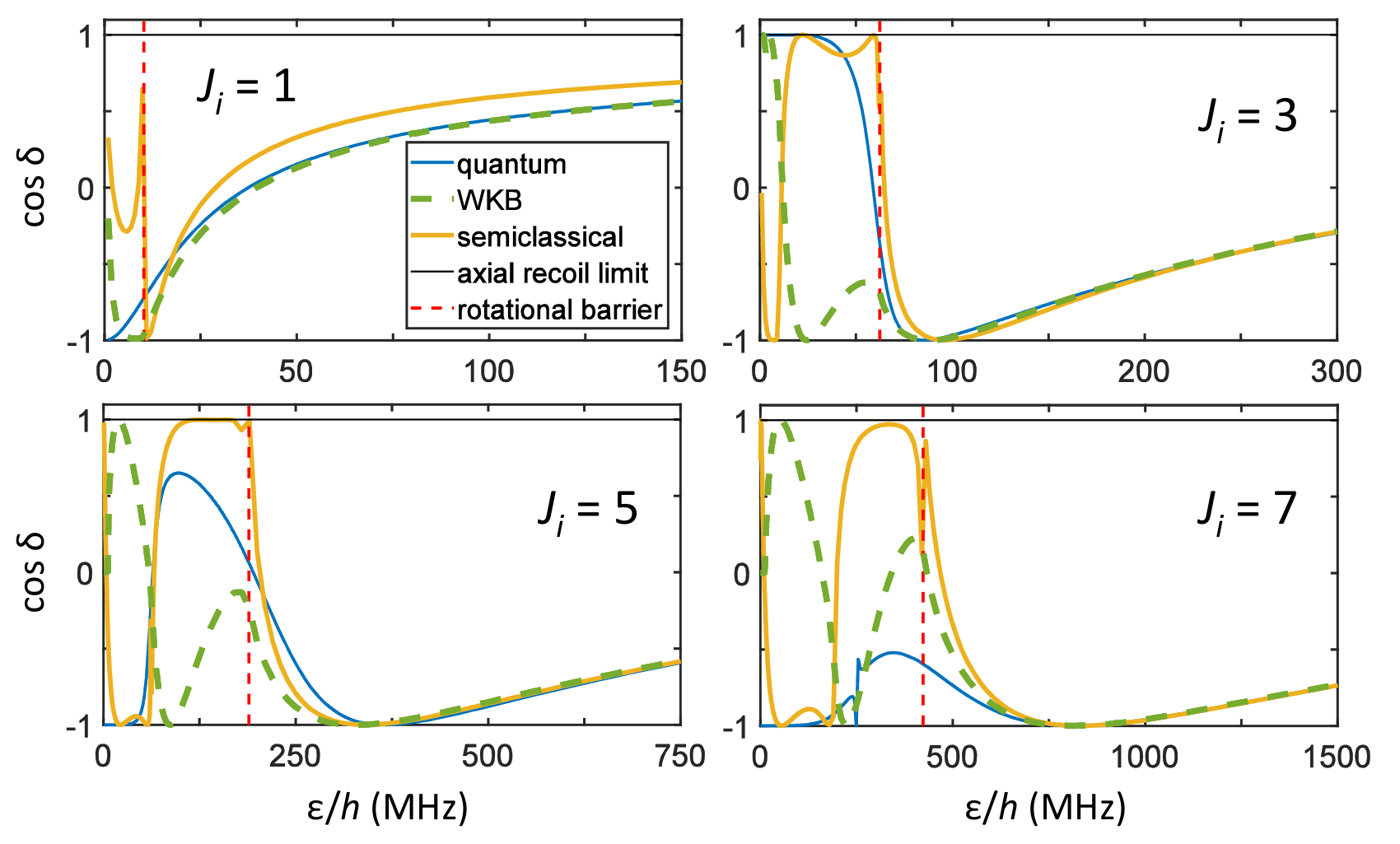}
\caption{A comparison of the quantum mechanical, WKB, and semiclassical methods for evaluating $\cos\delta$ for a range of initial molecular angular momenta $J_i$ within the $0_u^+$ electronic manifold (here $M_i=P=0$).  The phase angle $\delta$ quantifies interference between the two allowed dissociation channels and plays a key role in the observed photofragment angular distributions.  Both approximations approach the quantum mechanical result at increasingly higher energies that are set by the rotational barrier heights (dashed vertical lines).  All methods recover the axial recoil limit (horizontal lines) at high energies, but this convergence is slower for the $\delta$ parameter than for the $R$ parameter in Fig. \ref{fig:RParameterBarrierHeight}.}
\label{fig:QuantumWKBSemiDelta}
\end{figure}
}

\newcommand{\QuasiclassicalBreakdown}{
\begin{figure}
\includegraphics*[trim = 0in 0in 2.5in 0in, clip, width=3.375in]{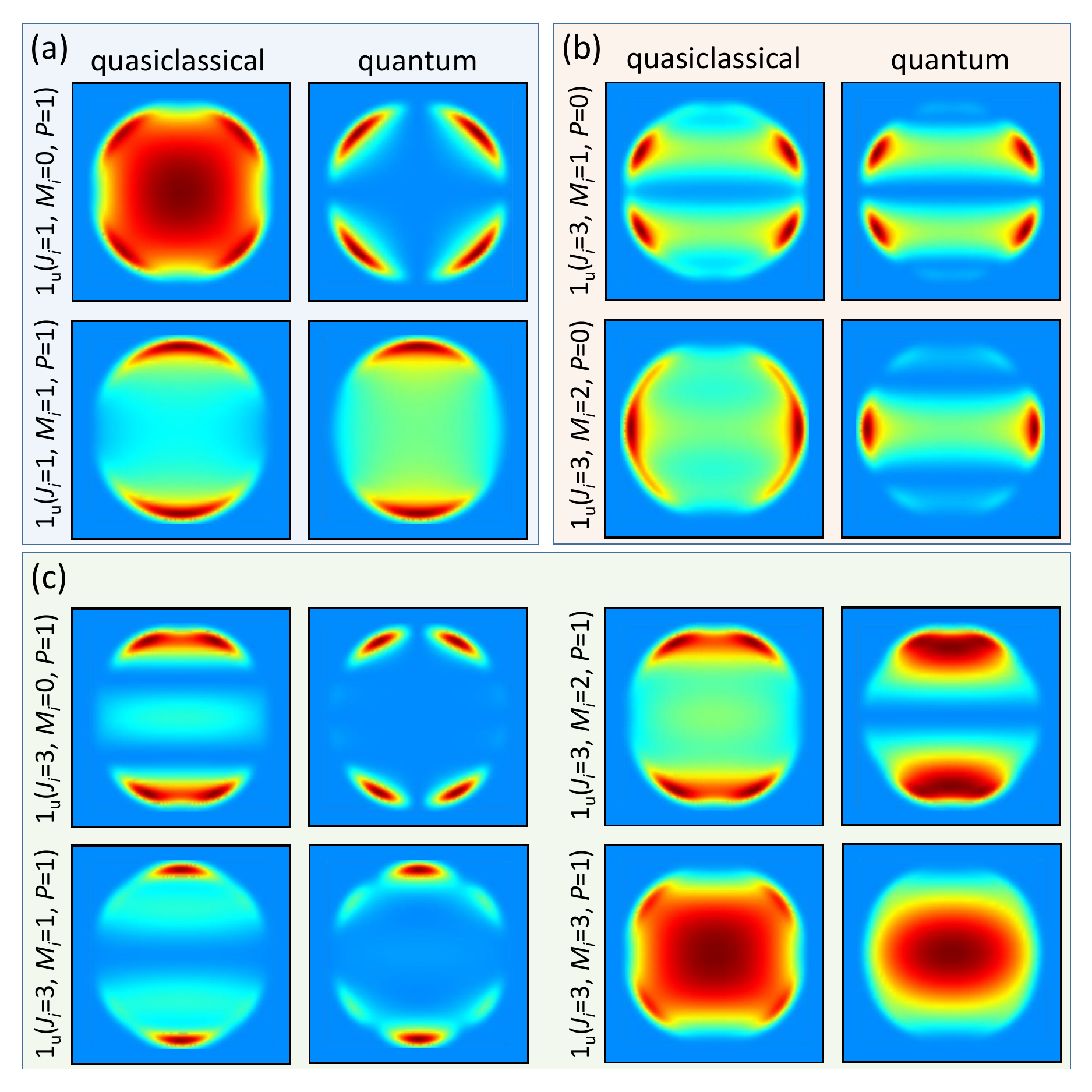}
\caption{Quasiclassical and quantum mechanical calculations of the photofragment angular distributions do not generally converge in the axial recoil limit if the photofragments are identical particles.  Experimental measurements always agree with the quantum theory \cite{ZelevinskyKondovArxiv18_QMQCPD}.  Here the high-energy angular distributions are calculated with the quasiclassical (\ref{eq:SigmaQuasiclassical}) and quantum mechanical (\ref{eq:SigmaAR}) models for the $1_u$ initial states with (a) $J_i=1,M_i=\{0,1\},P=1$; (b) $J_i=3,M_i=\{1,2\},P=0$; and (c) $J_i=3,M_i=\{0,1,2,3\},P=1$.}
\label{fig:QuasiclassicalBreakdown}
\end{figure}
}

\section{Introduction}
\label{sec:Introduction}

Low-temperature atomic and molecular collisions and reactions are qualitatively different from their high-energy counterparts.  While the latter can usually be explained quasiclassically, the former are strongly influenced by the wave nature of the particles and are correctly described only with quantum mechanical wave functions.  In addition, at very low temperatures, precise control of internal molecular states is possible, permitting studies of state-specific reaction cross sections rather than statistical averages.  Ultracold, quantum mechanical chemistry has many distinguishing features such as reaction barrier tunneling, quantum interference, and possibilities for controlling the outcome with applied electric or magnetic fields \cite{BalakrishnanJCP16_UltracoldMolecules}.  On the other hand, a quasiclassical interpretation can offer intuitive insight into complicated processes.

In particular, photodissociation \cite{SatoCR01_PhotodissociationReview,HerschbachZarePIEEE63_DiatomicPhotodissociation,ZareMPC72_PhotoejectionDynamics} has been extensively used to study the nature of molecular bonding, which becomes encoded in angular distributions of the outgoing photofragments.  It is one of the most basic chemical processes, and is highly amenable to quantum state control of the outgoing particles, since the reaction proceeds without a collision.  The majority of photodissociation experiments to date have been carried out in a regime that is well described in the quasiclassical framework \cite{BernsteinChoiJCP86_StateSelectedPhotofragmentation,ZareCPL89_PhotofragmentAngularDistributions,SeidemanCPL96_MagneticStateSelectedPDDistributions,ZareBeswickJCP08_PhotofragmentAngularDistrQuantClass}, while recently we studied this reaction in the purely quantum regime to observe matter-wave interference of the photodissociation products \cite{ZelevinskyMcDonaldNature16_Sr2PD} and control of the reaction by weak magnetic fields \cite{ZelevinskyMcDonaldPRL18_PDMagneticField}.  In this work, we show that the photodissociation energy can be tuned by several orders of magnitude to span the range from the quantum mechanical to the quasiclassical regime.  Experimentally and theoretically we demonstrate how the crossover occurs and what determines its energy scale.  We also find that if the photofragments are identical bosonic or fermionic particles, then the reaction outcome is not always well described in quasiclassical terms even at high energies and the effects of quantum statistics can persist indefinitely.  Moreover, we apply the Wentzel{-}Kramers{-}Brillouin (WKB) approximation to photodissociation, as well as a related semiclassical approximation, and discuss how accurately they capture the transition from ultracold to quasiclassical behavior.

In the experiment, we trap diatomic strontium molecules ($^{88}$Sr$_2$) at a temperature of a few microkelvin, enabling precise manipulation and preparation of specific quantum states as the starting point for photodissociation.  The photodissociation laser light has strictly controlled frequency and polarization, yielding photofragments in fully defined quantum states.  The \textit{ab initio} theory is aided by a state-of-the-art molecular model \cite{MoszynskiSkomorowskiJCP12_Sr2Dynamics,KillianBorkowskiPRA14_SrPAMassScaling} that captures the effects of nonadiabatic mixing and yields excellent agreement with spectroscopic measurements \cite{ZelevinskyMcGuyerPRL13_Sr2ZeemanNonadiabatic,ZelevinskyMcGuyerPRL15_Sr2ForbiddenE1,ZelevinskyMcGuyerNJP15_Sr2Spectroscopy,ZelevinskyMcGuyerNPhys15_Sr2M1}.

This approach offers a unique opportunity to finely scan a large range of energies that is relevant to the crossover from distinctly quantum mechanical to quasiclassical photodissociation.  Since initial molecular states have low angular momenta, the relevant dynamics occurs at very low, millikelvin energies.  The experiment can sample energies from $\sim0.1$ mK limited by the molecule trap depth to $\sim100$ mK limited by the available laser intensity that is needed to overcome the diminishing transition strengths.  The rotational and electronic potential barriers that are explored in this and related \cite{ZelevinskyKondovArxiv18_QMQCPD} work fall within this range, permitting direct observations of the crossover.  We reduce the process to a single initial molecular state and two angular-momentum product channels, in which case the quantum mechanics of the reaction can be encoded in a single pair of coefficients:  the amplitude and phase of the channel interference.

Note that in both quantum and quasiclassical regimes we work with single initial quantum states.  The classical nature of the process emerges not because of statistical averages over the initial states, but when the kinetic energies of the photofragments are large.  At high energies, the photodissociation time scale is much faster than the time scale of molecular rotations, and the molecules can be described as classical rotors.  An equivalent view is that the photodissociation outcome should have quasiclassical behavior if the photofragment energy in the continuum exceeds the height of any potential barriers.  This regime is referred to as axial recoil, where the molecule has insufficient time to rotate during photodissociation and the photofragments emerge along the instantaneous direction of the bond axis.  Here we investigate at what energies the axial recoil regime is reached, whether it is accurately described by a quasiclassical model, and how its onset depends on the molecular binding energy and angular momentum.

In Sec. \ref{sec:Experiment} of this work, we describe the photodissociation experiment and show the measured and calculated photofragment angular distributions.  In Sec. \ref{sec:PADs}, we discuss the quantum mechanical photodissociation cross section used in calculating the angular distributions, its axial recoil limit, and the quasiclassical model.  Section \ref{sec:Crossover} details the crossover from the ultracold to the quasiclassical regime for a range of molecular binding energies and angular momenta and shows the onset of the axial recoil limit at energies that exceed the potential barriers in the continuum.  In Sec. \ref{sec:QuantumStatistics} we discuss the breakdown of the quasiclassical picture of photodissociation in case of identical photofragments and present the correct high-energy limit that takes quantum statistics into account.  We summarize our conclusions in Sec. \ref{sec:Conclusions}.

\section{The photodissociation experiment}
\label{sec:Experiment}

The starting point for the experiment is creation of $\sim7,000$ weakly bound ultracold $^{88}$Sr$_2$ molecules in the electronic ground state.  This is accomplished by laser cooling, trapping, and photoassociating Sr atoms in a far-off-resonant optical lattice \cite{ZelevinskyReinaudiPRL12_Sr2}.  Subsequently, optical selection rules allow specific molecular quantum states to be populated and serve as a starting point for photodissociation.

\Schematic
Figure \ref{fig:Schematic} shows the photodissociation experiment from the point of view of molecular potentials and wave functions.  The ground state \textit{gerade} potential $X0_g^+$ correlates to the $^1S+{^1S}$ atomic threshold, while the singly-excited \textit{ungerade} potentials $0_u^+$ and $1_u$ correlate to $^1S+{^3P_1}$, where the 0 and 1 state labels correspond to $\Omega$, the projection of the total atomic angular momentum onto the molecular axis.  Weakly bound ground-state molecules are excited to $0_u^+$ and $1_u$ bound states with a laser pulse, and the resulting molecules are immediately photodissociated to the ground-state continuum.  The bound states are labeled by their vibrational quantum number $v$ (negative numbers count down from the threshold), total angular momentum $J$, and its projection $M$ onto the quantization axis that is set by a small vertical magnetic field.  In Fig. \ref{fig:Schematic}, the $0_u^+(v=-2,J_i=1,M_i=0)$ molecules are photodissociated by 689 nm light.  The allowed angular-momentum states, or partial waves, in the continuum are $J=\{0,2\}$ due to electric-dipole selection rules as well as the bosonic nature of $^{88}$Sr that allows only even $J$ values in the ground state.

The photodissociation laser light co-propagates with the lattice laser and both are focused to a $\sim30$ $\mu$m waist at the molecular cloud.  A broad imaging beam, resonant with a strong 461 nm transition in atomic Sr, nearly co-propagates with the lattice and is directed at a charge-coupled device camera that collects absorption images of the photofragments.  A typical experiment uses $\sim10$ $\mu$s long photodissociation laser pulses, $\sim100$ $\mu$s of free expansion, and $\sim10$ $\mu$s absorption imaging pulses \cite{ZelevinskyKondovArxiv18_QMQCPD}. Several hundred images are averaged in each experiment.

\PADs
The experimental and theoretical results of photodissociating the molecules in a range of precisely prepared quantum states are shown in Fig. \ref{fig:PADs}.  In this set of measurements, light polarization is along the quantization axis and the molecules start from $J_i=1$, $M_i=0$.  In Fig. \ref{fig:PADs}(a), molecules in vibrational states $v=\{-2,-3,-4,-5\}$ of $0_u^+$ are photodissociated.  The photofragment angular distributions show a strong energy dependence, and a less pronounced dependence on $v$ except for the most weakly bound levels, as discussed in Sec. \ref{sec:CrossoverR}.  The bottom row of theoretical images is obtained with the WKB approximation, which is independent of $v$ and described in Sec. \ref{sec:CrossoverDelta}.  Figure \ref{fig:PADs}(b) is analogous to (a) but illustrates $1_u$ initial states.  A good agreement between measurements and quantum mechanical calculations is reached for all initial states and continuum energies.  Figure \ref{fig:PADs}(c) shows agreement between quantum theory and the WKB approximation at the continuum energy $\varepsilon/h=1000$ MHz ($\varepsilon/k_B=48$ mK) where $h=2\pi\hbar$ and $k_B$ are the Planck and Boltzmann constants.  At this high energy, both methods approach the axial recoil limit which is also shown.

\section{Photofragment angular distributions}
\label{sec:PADs}

Sections \ref{sec:GeneralQModel}-\ref{sec:QuasiclassicalARA} overview the quantum mechanical calculation and parametrization of the photodissociation cross section both in the low-energy regime and in the high-energy axial recoil limit, as well as the correspondence of the latter to the quasiclassical approximation.

\subsection{General quantum mechanical model}
\label{sec:GeneralQModel}

The general quantum mechanical expression for the photodissociation cross section is based on Fermi's golden rule with the electric-dipole (E1) transition operator,
\begin{align}
\label{eq:SigmaQMGeneral}
& \sigma_{\mathrm{QM}}(\theta, \phi) \propto
 \\ \nonumber & \bigr\vert\left\langle \Psi_{\textbf{k}}^{JM}\left(\{\textbf{r}\},\mathcal{R}\right) \bigr\vert \hat{T}^1_{\mathrm{E1}} \bigr\vert \Psi_{J_iM_i\Omega_i} \left(\{\textbf{r}\}, \mathcal{R} \right) \right\rangle \bigr\vert^2,
\end{align}
where
$\{\textbf{r}\}$ is the set of atomic electronic coordinates,
$\mathcal{R}$ is the vector connecting the photofragments, $\textbf{k}$ is the lab-frame wave vector of the photofragments, $\Psi_{\textbf{k}}^{JM}$ is the final (continuum) wave function, and $\Psi_{J_iM_i\Omega_i}$ is the initial (bound) wave function.  The polar and azimuthal angles $\{\theta,\phi\}$ are referenced to the quantization axis.

Assuming a pure initial state $|J_iM_i \Omega_i \rangle$, expanding the bound and continuum wavefunctions into the products of their electronic, rovibrational and angular parts, integrating over the rotation angles and summing over $M$ transforms Eq. (\ref{eq:SigmaQMGeneral}) into
\begin{widetext}
\begin{align}
\label{eq:SigmaQM}
\sigma_{\mathrm{QM}}(\theta, \phi) \propto & \sum_{\Omega_k} \Bigr\rvert \sum_{J_k (M_k) P Q n_k} (-1)^{M_k + \Omega_k}(2 J_k+1)  D^{J_k}_{M_k \Omega_k} (\phi,\theta,0)
\\ \nonumber & \times
 \begin{pmatrix}
  J_k & 1 & J_i\\
  -M_k &P& M_i
 \end{pmatrix}
 \begin{pmatrix}
  J_k & 1 & J_i\\
  -\Omega_k &Q& \Omega_i
 \end{pmatrix}
\left\langle \chi_{n_k\Omega_k}^{J_k}(\mathcal{R})  \left| \mathbf{d}_{{\text{BF}}} \right| \chi_{n_iJ_i\Omega _i}(\mathcal{R}) \right\rangle \Bigr\rvert^2,
\end{align}
\end{widetext}
where $\chi_n(\mathcal{R})$ are the rovibrational wave functions indexed by all the relativistic electronic channels that are included in the model, $\mathbf{d}_{\mathrm{BF}}$ is the body-fixed E1 transition operator, $J_k$ are the indexed continuum angular momenta, $D_{M\Omega}^J$ are Wigner rotation matrices, and $Q=\Omega_k-\Omega_i$ (a photodissociation transition is called parallel if $Q=0$ and perpendicular if $|Q|=1$).  The polarization index $P=0$ or $P=\pm1$ if the photodissociation light is polarized along or perpendicularly to the quantization axis, respectively (such that $P=M_k-M_i$).

\subsection{Axial recoil approximation}
\label{sec:ARA}

At low photofragment energies, to obtain the correct photodissociation cross sections it is crucial to calculate the matrix elements $\langle \chi_{n_k\Omega_k}^{J_k}(\mathcal{R})  | \mathbf{d}_{{\text{BF}}}  | \chi_{n_iJ_i \Omega _i}(\mathcal{R})  \rangle$ in Eq. (\ref{eq:SigmaQM}). At high energies, these matrix elements become independent of $J_k$ \cite{SeidemanCPL96_MagneticStateSelectedPDDistributions} and can be simply factored out of Eq. (\ref{eq:SigmaQM}).  This is equivalent to disregarding the photodissociation dynamics.  Under this axial recoil approximation, the photodissociation cross section (\ref{eq:SigmaQM}) reduces to
\begin{widetext}
\begin{align}
\label{eq:SigmaAR}
 \sigma(\theta, \phi)_{\mathrm{AR}} \propto & \sum_{Q} \Biggr\rvert\sum_{J_k P } (-1)^{M_k + \Omega_k}(2 J_k+1)  D^{J_k}_{M_k \Omega_k} (\phi,\theta,0)
 \begin{pmatrix}
  J_k & 1 & J_i\\
  -M_k &P& M_i
 \end{pmatrix}
 \begin{pmatrix}
  J_k & 1 & J_i\\
  -\Omega_k &Q& \Omega_i
 \end{pmatrix}\Biggr\rvert^2.
\end{align}
\end{widetext}

\subsection{Two-parameter quantum mechanical model}
\label{sec:TwoParameterModel}

Consider the photodissociation of $0_u^+(v,J_i,M_i)$ molecules to the ground-state continuum with $P=0$.  For odd $J_i$ (which are the only allowed angular momenta in $0_u^+$ due to quantum statistics), the allowed angular momenta in the continuum are $J=J_i-1$ and $J_i+1$, while $M=M_i$.
Then the quantum mechanical photofragmentation cross section can be expressed using only two parameters,
the channel amplitude-squared $R$ and the relative channel phase $\delta$:
\begin{align}
\label{eq:IQM2J}
& \sigma_{R\delta}(\theta,\phi) \propto
\\ \nonumber & \left|\sqrt{R} Y_{J_i-1, M_i}(\theta,\phi) + (-1)^{\Omega_i} e^{i \delta} \sqrt{1-R} Y_{J_i+1, M_i}(\theta,\phi)\right|^2.
\end{align}
The $(-1)^{\Omega_i}$ factor correctly connects the sign of the $\delta$ parameter to the phase shift between the continuum wave functions, and $Y_{JM}$ are spherical harmonics which are proportional to the angular wave functions for $\Omega=0$.

The expression for the $R$ parameter can be derived by comparing Eq. (\ref{eq:IQM2J}) to the traditional cross section (\ref{eq:SigmaQM}).  We obtain
\begin{widetext}
\begin{align}
\label{eq:RQuantum}
 R & = \left| \sqrt{2 J_i - 1}
 \begin{pmatrix}
  J_i - 1 & 1 & J_i\\
  -M_k &P& M_i
 \end{pmatrix}
 \begin{pmatrix}
  J_i - 1 & 1 & J_i\\
  0 &-\Omega_i& \Omega_i
 \end{pmatrix}
 \left\langle \chi_{n_k\Omega_k}^{J=J_{i-1}}(\mathcal{R}) \left| \mathbf{d}_{{\text{BF}}} \right| \chi_{n_iJ_i \Omega _i}(\mathcal{R})  \right\rangle
 \right|^2
 \\ \nonumber
 & \times \left[ \sum_{J_k = \{ J_i - 1, J_i + 1 \}} \left| \sqrt{2 J_k + 1}   \begin{pmatrix}
  J_k & 1 & J_i\\
  -M_k &P& M_i
 \end{pmatrix}
 \begin{pmatrix}
  J_k & 1 & J_i\\
  -\Omega_k &Q& \Omega_i
 \end{pmatrix}
\left\langle \chi_{n_k\Omega_k}^{J_k}(\mathcal{R}) \left| \mathbf{d}_{{\text{BF}}} \right| \chi_{n_iJ_i \Omega _i}(\mathcal{R}) \right\rangle\right|^2 \right]^{-1}.
\end{align}
\end{widetext}
The $ \delta$ parameter is calculated as the phase shift difference for the continuum wave functions, $\delta = \delta_{J_i + 1} - \delta_{J_i - 1}$.

In the axial recoil limit Eq. (\ref{eq:RQuantum}) is simplified by setting the matrix elements to a constant and canceling them, as explained in Sec. \ref{sec:ARA}.
This results in values of $R$ that approach 0.5 with increasing $J_i$, while the phase shifts become identical:  $\delta_{J_i + 1} \rightarrow \delta_{J_i - 1}$, yielding $\delta \rightarrow 0$ and $\cos \delta \rightarrow 1$.

\subsection{Quasiclassical model}
\label{sec:QuasiclassicalModel}

The commonly used quasiclassical model for photofragment angular distributions \cite{BernsteinChoiJCP86_StateSelectedPhotofragmentation,ZareCPL89_PhotofragmentAngularDistributions} describes the photodissociation cross section as
\begin{align}
\label{eq:SigmaQuasiclassical}
  \sigma_{\mathrm{QC}}(\theta, \phi) \propto  P_i(\theta, \phi) \times
  \left[1 + \beta_2 P_2 (\cos \theta)\right],
\end{align}
where $P_i(\theta, \phi) = |D^{J_i}_{M_i \Omega_i}(\phi, \theta, 0)|^2$ is the probability density of the initial molecular orientation.  The anisotropy parameter is $\beta_2=2$ for parallel photodissociation transitions ($Q=0$) and $\beta_2 = -1$ for perpendicular transitions ($|Q|= 1$).
If the molecule is in a superposition of $| J_i M_i \Omega_i \rangle$ states, the initial probability density can be generalized as \cite{ZareBeswickJCP08_PhotofragmentAngularDistrQuantClass}
\begin{align}
\label{eq:InitialDensityGen}
P_i(\theta, \phi) = \sum_{J_i J_i' M_i M_i' \Omega_i } D^{J_i}_{M_i \Omega_i}(\phi, \theta, 0) D^{J_i' \star}_{M_i' \Omega_i}(\phi, \theta, 0).
\end{align}
Note that in Eq. (\ref{eq:InitialDensityGen}) there are no cross terms for $\Omega_i$.  The intuition behind Eq. (\ref{eq:SigmaQuasiclassical}) is that the photofragment angular distribution follows the initial angular density of the molecule, modified by the angular probability density of the photon absorption.

\subsection{Correspondence of the quasiclassical model to the axial recoil approximation}
\label{sec:QuasiclassicalARA}

It was shown in Ref. \cite{SeidemanCPL96_MagneticStateSelectedPDDistributions} that for photodissociation light polarized along the quantization axis, Eq. (\ref{eq:SigmaAR}) reduces to $\sigma_{\mathrm{AR}}(\theta,\phi)\propto|D_{M_i\Omega_i}^{J_i}|^2\times[1+\beta_2P_2(\cos\theta)]$ for initial states with a single $\Omega_i$ \cite{ZareBeswickJCP08_PhotofragmentAngularDistrQuantClass}.
We have confirmed that this result also holds for other polarizations of the photodissociating light (if the light polarization is perpendicular to the quantization axis, $P_2(\cos\theta)$ should be replaced with $P_2(\sin\theta\sin\phi)$).

Therefore the quasiclassical model (\ref{eq:SigmaQuasiclassical}) can be applied in the axial recoil limit to obtain angular distribution predictions that are identical to those of the quantum mechanical model.  However, an important assumption in Refs. \cite{SeidemanCPL96_MagneticStateSelectedPDDistributions,ZareBeswickJCP08_PhotofragmentAngularDistrQuantClass} is that the photofragments are not identical quantum particles.  Deviations from this assumption are discussed in Sec. \ref{sec:QuantumStatistics}.

\section{Crossover from the ultracold regime to the axial recoil limit}
\label{sec:Crossover}

We investigate the crossover from the ultracold quantum mechanical to the high-energy axial recoil regime of photodissociation, as observed in Fig. \ref{fig:PADs}.  The intuitive parametrization of Eq. (\ref{eq:IQM2J}) that accurately illustrates two-channel photodissociation allows us to study the crossover as it applies to only two parameters, $R$ and $\delta$.  These parameters can have a strong energy dependence in the ultracold regime while approaching their axial recoil values at higher energies.  In this Section, we address the following questions:  at what energy scale does the outcome of photodissociation approach the axial recoil limit; how does this energy scale depend on the molecular binding energy (or vibrational level) and rotational angular momentum; and how well do the WKB and semiclassical approximations model this crossover?

\subsection{Approaching the axial recoil limit:  Channel amplitude}
\label{sec:CrossoverR}

\RParametersV
The $R$ parameter from Eq. (\ref{eq:IQM2J}) depends on both the bound and continuum wave functions, as can be explicitly seen from Eq. (\ref{eq:RQuantum}).  Therefore, its behavior as a function of continuum energy can depend on the vibrational state of the molecule, potentially influencing the energy scale at which the axial recoil limit is reached based on the molecular binding energy.
The plots in Fig. \ref{fig:RParametersV}(a) show the $R$ parameter for several weakly bound initial states $0_u^+(v,J_i=1,M_i=0)$ with $v=\{-2,-3,-4,-5\}$ as well as for the most deeply bound state $v=0$.  The dashed line indicates the axial recoil limit.

For some of the shallow vibrational levels such as $v=-2,-3$ (bound by $\sim10^2$ MHz) the $R$ parameter shows strong energy dependence that does not closely resemble that of more deeply bound states, as is also observed in Fig. \ref{fig:PADs}.  This is caused by the large spatial extent ($\gtrsim100$ bohr) of weakly bound wave functions, as illustrated for the $0_u^+(-2,1,0)$ wave function in Fig. \ref{fig:Schematic}.  The continuum wave functions, as also shown in Fig. \ref{fig:Schematic}, are similar at small internuclear separations but undergo a relative phase shift at larger distances that correspond to the bond length in initially very weakly bound molecules.  This can prevent photodissociation from reaching quasiclassical behavior at the expected energies.
The energy dependences for $v=-4,-5$ (bound by $\sim10^3$ MHz) are more regular, and we find that all the deeper vibrational levels starting from $v=-10$ (bound by $\sim10^5$ MHz) exhibit nearly identical behavior illustrated in Fig. \ref{fig:RParametersV}(a) by the deepest $v=0$ level (bound by $\sim10^8$ MHz).  We conclude that with the exception of asymptotic molecules that are bound just below threshold, the $R$ parameter is independent of the vibrational quantum number.  Since the $\delta$ parameter is $v$-independent as discussed in Sec. \ref{sec:CrossoverDelta}, the expected anisotropy patterns are nearly independent of the binding energy, or $v$.  We have checked that for higher initial angular momenta $J_i$, the $R$ parameter is also independent of $v$ except for the most weakly bound molecules.

It is clear from Fig. \ref{fig:RParametersV}(a) that $R\rightarrow1$ as $\varepsilon\rightarrow0$, indicating that the lowest partial wave ($J=0$ in this case) dominates the photofragment distribution at ultralow energies.  This is caused by the suppression of low-energy scattering wave functions with higher angular momenta at relevant internuclear separations, due to their rotational barriers.  Figure \ref{fig:RParametersV}(b) shows the $J=0$ and $J=2$ wave functions at the low continuum energy of $\varepsilon/h=0.5$ MHz ($\varepsilon/k_B=24$ $\mu$K), where $J=2$ is strongly suppressed at the relevant molecular bond lengths of $\lesssim100$ bohr.
\vspace*{7px}

\RParameterBarrierHeight
The crossover from the ultracold, quantum mechanical regime to the axial recoil limit is explored in Fig. \ref{fig:RParameterBarrierHeight} as a function of the initial angular momentum $J_i$.  The $R$ parameter is plotted as a function of energy for different odd $J_i$ of initial states within the $0_u^+$ and $1_u$ electronic manifolds, assuming $M_i=P=0$.  The calculations were performed for the most deeply bound vibrational level $v=0$ to avoid any dependence on $v$.  For all initial states, $R=1$ at threshold and decays to the axial recoil limit (shown with horizontal lines) as a function of energy.
In each frame, a vertical line marks the rotational barrier height corresponding to the largest partial wave in the continuum, $J_i+1$.  As expected, the barrier height sets the energy scale for the onset of the axial recoil regime.  This trend is evident despite the irregularities connected to shape resonances for $J=\{4,8,20\}$ (the $J=4$ shape resonance has been experimentally detected \cite{ZelevinskyMcDonaldNature16_Sr2PD}).

\subsection{Approaching the axial recoil limit:  Channel phase and various approximations}
\label{sec:CrossoverDelta}

The $\delta$ parameter in Eq. (\ref{eq:IQM2J}) is the phase difference between the continuum channel wave functions and is independent of the initial state of the photodissociated molecules.  It is thus readily amenable to a range of approximation techniques.  We consider the WKB approximation \cite{ChildBook,NikitinUmanskiiBook} and the semiclassical approximation \cite{AshfoldWredeJCP02_AxialRecoilBreakdown}, and investigate their applicability to photodissociation across a wide span of energies as well as their convergence to the axial recoil limit.

The WKB approximation is a method of solving linear differential equations with spatially varying coefficients.  It is applicable to approximating the solutions to the time-independent Schr\"{o}dinger equation, where the wave function is represented as an exponential with smoothly varying amplitude and phase, and is a particularly useful method of finding $\delta$ without using the full quantum mechanical treatment.

The continuum wave function corresponding to the partial wave $J$ has the asymptotic behavior
\begin{align}
\label{eq:PsiAsymptotic}
 \psi_{J}(\mathcal{R}) \propto \sin \left(k\mathcal{R} + \delta_{J} - \frac{J \pi}{2} \right),
\end{align}
where $k\equiv\sqrt{2 \mu\varepsilon}/\hbar$ and $\mu$ is the reduced mass.
An analytical expression for $\delta_{J}$ is based on the WKB approximation where the wave function is assumed to have the form $\psi_J(\mathcal{R})=e^{iS(\mathcal{R})/\hbar}$.  If only the lowest-order term in $\hbar$ is kept in $S(\mathcal{R})$, then $\psi(\mathcal{R})\propto\sin(\int_{\mathcal{R}_0}^\mathcal{R}p(\mathcal{R})d\mathcal{R}/\hbar+\pi/4)$ where $\mathcal{R}_0$ is the classical turning point and $p(\mathcal{R})\equiv\sqrt{2\mu(E-V(\mathcal{R}))-J(J+1)/\mathcal{R}^2}$ in terms of the molecular potential $V(\mathcal{R})$.  The WKB approximation is most applicable at higher energies and breaks down near the turning point where $p(\mathcal{R})\approx0$.  Based on the WKB expression for $\psi(\mathcal{R})$ and Eq. (\ref{eq:PsiAsymptotic}), the asymptotic phase shift is
\begin{align}
\label{eq:WKBDelta}
\delta_J^{\mathrm{WKB}} = \lim_{\mathcal{R} \rightarrow \infty} \left[ \frac{1}{\hbar}\int_{\mathcal{R}_0}^{\mathcal{R}}  p(\mathcal{R}) d\mathcal{R} + \frac{\pi}{4} - k\mathcal{R} +\frac{J \pi}{2} \right].
\end{align}

A related approach to investigating the quantum-quasiclassical crossover is to consider the classical rotation of the molecule during photodissociation.  We refer to this method as the semiclassical approximation.  Its value is in the intuitive interpretation of the molecular rotation angle during the bond breaking process and a slightly faster computational convergence than the WKB method.  The angle of rotation $\gamma$ between the initial molecular axis and the axis created by the scattered fragments is $\gamma_J=\int_{\mathcal{R}_0}^{\infty}\hbar\sqrt{J(J+1)}/[p(\mathcal{R})\mathcal{R}^2]d\mathcal{R}$.
This angle is connected to the anisotropy parameter $\beta_2$ by the relation $\beta_2 = \beta_{\mathrm{AR}} P_2(\cos\gamma)$,
where $\beta_{\mathrm{AR}}$ takes on the limiting values from the quasiclassical model \cite{ZareMPC72_PhotoejectionDynamics,BernsteinChoiJCP86_StateSelectedPhotofragmentation,ZareCPL89_PhotofragmentAngularDistributions} such that $\beta_{\mathrm{AR}}=\{2,-1\}$ for parallel and perpendicular photodissociation transitions, respectively.  This semiclassical approximation is more general than the quasiclassical model which assumes that $\gamma=0$.\

The WKB method links the classical parameter $\gamma$ to the quantum mechanical phase $\delta_J$ through $\gamma=-d\delta_J/dJ$ \cite{NikitinUmanskiiBook}.  Further approximating the derivative by a difference quotient yields the semiclassical approximation for the phase shifts,
\begin{align}
\label{eq:SemiclassicalDelta}
  \delta_J-\delta_{J'} = (J'-J) \frac{\gamma_J + \gamma_{J'}}{2}.
\end{align}

The semiclassical approximation has been successfully applied to diatomic molecule photodissociation at the energy of several wave numbers above threshold \cite{AshfoldWredeJCP02_AxialRecoilBreakdown}, where a deviation from the axial recoil limit was observed, yielding an agreement with quantum mechanical calculations \footnote{In Ref. \cite{AshfoldWredeJCP02_AxialRecoilBreakdown} the semiclassical model is referred to as ``quasiclassical''.  However, it is not equivalent to the quasiclassical description from Refs. \cite{BernsteinChoiJCP86_StateSelectedPhotofragmentation,ZareCPL89_PhotofragmentAngularDistributions}, and is a more accurate approximation.}.
To summarize, the hierarchy of approximations, from the most to the least exact, is (i) quantum mechanical treatment, (ii) the WKB approximation, (iii) the semiclassical model, and (iv) the axial recoil approximation (which is generally equivalent to the quasiclassical model with exceptions discussed in Sec. \ref{sec:QuantumStatistics} and Ref. \cite{ZareBeswickJCP08_PhotofragmentAngularDistrQuantClass}).

\QuantumWKBSemiDelta
Figure \ref{fig:QuantumWKBSemiDelta} shows a comparison of the quantum mechanical (Eq. (\ref{eq:SigmaQM})), WKB (Eq. (\ref{eq:WKBDelta})), and semiclassical (Eq. (\ref{eq:SemiclassicalDelta})) approaches to calculating the $\delta$ parameter.  This phase angle is determined for the initial molecular states $J_i=\{1,3,5,7\}$ within the $0_u^+$ electronic manifold, assuming photodissociation with $M_i=P=0$.  We find the applicability ranges of both approximation methods to be similar, with the exception that the WKB approximation is more accurate for very low $J_i$.  These ranges are set by the barrier heights in the continuum, indicated with dashed vertical lines.  We also find that all methods converge to the axial recoil limit $\cos\delta=1$.  However, this convergence happens more slowly than for the $R$ parameter in Fig. \ref{fig:RParameterBarrierHeight}, and dominates the discrepancy between the measured angular distributions and those expected for axial recoil at energies beyond the barrier height, as in Fig. \ref{fig:PADs} where the barrier height is only $\sim10$ MHz but nonclassical behavior persists to $\sim10^3$ MHz \footnote{For the WKB images in Fig. \ref{fig:PADs}, the axial-recoil value of $R$ was used, since $R$ approaches the high-energy limit faster than $\delta$.}.  Furthermore, only the quantum mechanical approach is sensitive to shape resonances, for example one that is visible in Fig. \ref{fig:QuantumWKBSemiDelta} at $\varepsilon/h\approx250$ MHz for $J_i=7$.

\section{The role of quantum statistics in photodissociation}
\label{sec:QuantumStatistics}

The equivalence of the quasiclassical approximation to the axial recoil limit of the quantum mechanical model holds for molecules in well-defined $\Omega_i$ states under the assumption that the photofragments are not identical bosons or fermions.  Section \ref{sec:QuasiclassicalARA} assumes that a molecule with angular momentum $J_i\neq0$ can dissociate into the $\{J_i-1,J_i,J_i+1\}$ partial waves as allowed by E1 selection rules.

Spin statistics can impose additional limitations on the available continuum channels, as is the case for Sr$_2$ and many other molecules.  In the electronic ground state, $J$ is always even for Sr$_2$ composed of identical bosonic Sr atoms and odd for Sr$_2$ composed of identical fermions.  In this work, we use bosonic $^{88}$Sr and therefore $J$ is even.  This restriction only affects $\Omega_k=0$ electronic states.  Furthermore, in cases where $\Omega_i=\Omega_k=0$ or $M_i=M_k=P=0$, the properties of the $3j$ symbols in Eqs. (\ref{eq:SigmaQM},\ref{eq:SigmaAR}) naturally force the continuum $J$ to be only even or odd, making the axial recoil approximation insensitive to quantum statistics \footnote{For $M_i=P=0$, quantum statistics (when combined with the properties of the $3j$ symbol in Eqs. (\ref{eq:SigmaQM},\ref{eq:SigmaAR})) can cause a normally allowed photodissociation pathway to become forbidden, for example for initial states $1_u(v,J_i,0)$ with even $J_i$ in bosonic-Sr dimers.  Such transitions, however, can be enabled with weak magnetic fields \cite{ZelevinskyMcDonaldNature16_Sr2PD,ZelevinskyMcDonaldPRL18_PDMagneticField}.}.  Here we extend Eq. (\ref{eq:SigmaAR}) to cases where quantum statistics must be considered, and compare the results to the quasiclassical model (\ref{eq:SigmaQuasiclassical}).  In our experiment, this affects photodissociation from the $1_u$ states to the ground-state continuum, except in cases where $M_i=P=0$.

First we assume that the photodissociation light polarization points along the quantization axis.  The two relevant cases are $J = J_i$ (applicable, for example, to fermionic-Sr dimers dissociating from odd $J_i$) and $J = \{ J_i-1, J_i+1 \}$ (applicable to bosonic-Sr dimers photodissociating from odd $J_i$ as in our experiments).  As shown in the Appendix, we adapt the axial-recoil cross section $\sigma_{\mathrm{AR}}(\theta,\phi)$ to both quantum-statistics restricted cases, and find the modified photodissociation cross section, $\sigma_{\mathrm{QS},\parallel}(\theta, \phi)$.  The result is analogous to the quasiclassical cross section $\sigma_{\mathrm{QC}}(\theta, \phi)$ in Eq. (\ref{eq:SigmaQuasiclassical}) but with the initial angular probability density $P_i(\theta, \phi)= |D^{J_i}_{M_i \Omega_i}(\phi, \theta, 0)|^2$ replaced by
\begin{align}
\label{eq:PParallel}
P_{i,\parallel}'(\theta, \phi) = \bigr\vert D^{J_i}_{M_i, 1 } (\phi,\theta,0) -\sigma_i D^{J_i}_{M_i , -1} (\phi,\theta,0)\bigr\vert^2,
\end{align}
where $\sigma_i\equiv p_i(-1)^{J_i}$ is the spectroscopic parity of the initial state and $p_i$ is the parity with respect to space-fixed inversion.  For example, for $1_u$ initial states, $p_i=-1$ for bosonic-Sr dimers and $p_i=1$ for fermionic-Sr dimers.

Next we treat the case where the photodissociation light polarization is perpendicular to the quantization axis.  Again spin statistics enforces either $J = J_i$ or $J = \{ J_i-1, J_i+1 \}$.  As detailed in the Appendix, we adapt $\sigma_{\mathrm{AR}}(\theta,\phi)$ to both cases and find the modified photodissociation cross section
\begin{align}
\label{eq:PPerp}
\sigma_{\mathrm{QS},\perp}(\theta, \phi) & \propto
\\ \nonumber\cos^2\phi & \bigr\vert D^{J_i}_{M_i, 1 } (\phi,\theta,0) +\sigma_i D^{J_i}_{M_i , -1} (\phi,\theta,0)\bigr\vert^2+
\\ \nonumber \cos^2\theta\sin^2\phi & \bigr\vert D^{J_i}_{M_i, 1 } (\phi,\theta,0) -\sigma_i D^{J_i}_{M_i , -1} (\phi,\theta,0)\bigr\vert^2.
\end{align}
This result is not analogous to the quasiclassical cross section $\sigma_{\mathrm{QC}}(\theta, \phi)$.

The photodissociation cross sections modified by spin statistics are not correctly described by the quasiclassical model (\ref{eq:SigmaQuasiclassical}), even in the axial recoil limit of large continuum energies.  This effect is observable in experiments as a deviation of high-energy photodissociation images from the quasiclassical model for certain initial molecular states and light polarizations \cite{ZelevinskyKondovArxiv18_QMQCPD}.

\QuasiclassicalBreakdown
Figure \ref{fig:QuasiclassicalBreakdown} illustrates $^{88}$Sr$_2$ photodissociation pathways where photofragment angular distributions in the axial recoil limit (\ref{eq:SigmaAR}) cannot be described by the quasiclassical model (\ref{eq:SigmaQuasiclassical}).
In our experiment and in the figure, the initial molecular states and laser polarizations that lead to a non-quasiclassical axial recoil limit are (a) $1_u(J_i=1,M_i=\{0,1\},P=1)$; (b) $1_u(J_i=3,\,M_i=\{1,2\},\,P=0)$; and (c) $1_u(J_i=3,\,M_i=\{0,1,2,3\},\,P=1)$.  The case considered in the top row of Fig. \ref{fig:QuasiclassicalBreakdown}(a) results in a distribution that does not evolve with energy, since only a single partial wave $J=2$ is allowed in the continuum by selection rules and hence the matrix elements in Eq. (\ref{eq:RQuantum}) cancel for all energies \footnote{Two other configurations with this property, $1_u(J_i=1,M_i=1,P=0)$ and $1_u(J_i=3,M_i=3,P=0)$, are not shown.}.  Our experimental measurements of angular distributions always agree with quantum mechanical calculations \cite{ZelevinskyKondovArxiv18_QMQCPD}.  For photodissociation of molecules in analogous quantum states but composed of identical fermionic isotopes we expect to observe different distributions than in Fig. \ref{fig:QuasiclassicalBreakdown}, and still different results, that are well described by the quasiclassical model, are expected for mixed dimers.

\section{Conclusions}
\label{sec:Conclusions}

We report photodissociation with ultracold $^{88}$Sr$_2$ molecules in isolated internal quantum states, including different binding energies and angular momenta, where we image and calculate the photofragment angular distributions as a function of the kinetic energy in the continuum, spanning from the ultracold to the quasiclassical regime.  At the higher energies that exceed the relevant potential barriers, the distributions converge to a small set of angular patterns that correspond to the axial recoil limit where the molecule has no time to rotate during the bond breaking process.  In contrast, at the lower energies the distributions exhibit strong variation with energy and a dependence on the initial molecular state.

We utilize precise quantum state selection to ensure a two-channel photodissociation outcome that can be described by only two parameters:  an amplitude that has a weak dependence on the initial vibrational quantum number and a phase that only depends on the continuum wave functions.  The amplitude converges to the axial recoil limit significantly faster than the phase, with a slower convergence for very weakly bound initial molecular states.  The WKB and semiclassical approximations are shown to agree with the quantum mechanical model at energies that exceed the barrier heights, and to correctly approach the axial recoil limit.  Finally, we find that in case of identical photofragments, the effects of bosonic or fermionic statistics can persist into the high energy regime, in which case the axial recoil limit disagrees with the ubiquitous quasiclassical model.

The ability to access the crossover from the ultracold to the quasiclassical regime of photodissociation was enabled by quantum state control of the molecules, making it possible to populate exclusively low-angular-momentum states and thus access low partial waves in the continuum that are strongly sensitive to quantum effects.  As more experiments begin to probe cold and ultracold chemistry, including molecular photodissociation, it is important to recognize the extent of the parameter space where the processes are truly quantum mechanical.  Our work brings this insight into a long-debated issue of the applicability of quasiclassical models to molecular photodissociation, and elucidates which parameters limit the convergence toward the high-energy limit.

\begin{acknowledgments}
We acknowledge the ONR Grant No. N00014-17-1-2246 and the NSF Grant No. PHY-1349725.  R. M. and I. M. also acknowledge the Polish National Science Center Grant No. 2016/20/W/ST4/00314 and M. M. the NSF IGERT Grant No. DGE-1069240.
\end{acknowledgments}

\appendix*
\section{Axial recoil limit for identical photofragments}
\subsection{Light polarization along the quantization axis}

With the assumptions $\{\Omega_i = 1,\Omega_k=0,M_k = M_i\}$, as for Sr$_2$ photodissociation from the $1_u$ states to the ground state continuum with light polarization along the quantization axis ($P=0$), the axial-recoil photodissociation cross section (\ref{eq:SigmaAR}) becomes
\begin{widetext}
\begin{align}
\label{eq:SigmaARAdapted1}
 \sigma(\theta, \phi)\propto & \Bigr\vert\sum_{J_k  } (-1)^{M_i }(2 J_k+1)  D^{J_k}_{M_i 0} (\phi,\theta,0)
 \begin{pmatrix}
  J_k & 1 & J_i\\
  -M_i &0& M_i
 \end{pmatrix}
 \begin{pmatrix}
  J_k & 1 & J_i\\
  0 &-1& 1
 \end{pmatrix}\Bigr\vert^2.
\end{align}
\end{widetext}

\subsubsection{Partial waves restricted to $J_i$}

After restricting the allowed partial waves to $J_k=J_i$ and transforming the D-matrix as
\begin{widetext}
\begin{align}
\label{eq:DMatrix}
D^{J_i}_{M_i 0} (\phi,\theta,0) = -\frac{\sqrt{J_i(J_i+1)}}{2M_i} \sin \theta \left[D^{J_i}_{M_i, -1 } (\phi,\theta,0) + D^{J_i}_{M_i , 1} (\phi,\theta,0)\right],
\end{align}
\end{widetext}
Eq. (\ref{eq:SigmaARAdapted1}) reads
\begin{align}
\sigma(\theta, \phi) \propto \bigr\vert D^{J_i}_{M_i, -1 } (\phi,\theta,0) + D^{J_i}_{M_i , 1} (\phi,\theta,0) \big\vert^2 \sin \theta^2.
\end{align}
This cross section has the form of Eq. (\ref{eq:SigmaQuasiclassical}) with $\beta_2=-1$ (as expected for this $\Delta\Omega=1$ transition), with the initial angular probability density replaced by the expression in Eq. (\ref{eq:PParallel}).

\subsubsection{Partial waves restricted to $J_i\pm1$}

After restricting the partial waves to $J_k=J_i\pm1$, the cross section in Eq. (\ref{eq:SigmaARAdapted1}) can be manipulated by taking the sum over all partial waves and subtracting the forbidden contributions,
\begin{widetext}
\begin{align}
\label{eq:SigmaDerivation1}
 \sigma(\theta, \phi)\propto & \Bigr\vert\sum_{J_k = \{ J_i-1,J_i,  J_i+1 \} } (-1)^{M_i }(2 J_k+1)  D^{J_k}_{M_i 0} (\phi,\theta,0)
 \begin{pmatrix}
  J_k & 1 & J_i\\
  -M_i &0& M_i
 \end{pmatrix}
 \begin{pmatrix}
  J_k & 1 & J_i\\
  0 &-1& 1
 \end{pmatrix} +
\\ \nonumber
 & - (-1)^{M_i }(2 J_i+1)  D^{J_i}_{M_i 0} (\phi,\theta,0)
 \begin{pmatrix}
  J_i & 1 & J_i\\
  -M_i &0& M_i
 \end{pmatrix}
 \begin{pmatrix}
  J_i & 1 & J_i\\
  0 &-1& 1
 \end{pmatrix}\Bigr\vert^2 =
 \\ \nonumber
 = & \Bigr\vert D^{1}_{0, -1}(\phi, \theta, 0) D^{J_i}_{M_i, 1}(\phi, \theta, 0) +  (-1)^{M_i}(2 J_i+1) \left[D^{J_i}_{M_i, -1 }  (\phi,\theta,0) + D^{J_i}_{M_i , 1} (\phi,\theta,0)\right] \sin \theta
 \\ \nonumber
  & \times \frac{\sqrt{J_i(J_i+1)}}{2M_i}
 \begin{pmatrix}
  J_i & 1 & J_i\\
  -M_i &0& M_i
 \end{pmatrix}
 \begin{pmatrix}
  J_i & 1 & J_i\\
  0 &-1& 1
 \end{pmatrix} \Bigr\vert^2 =
 \\ \nonumber
  = & \Bigr\vert- \frac{1}{\sqrt{2}} \sin \theta D^{J_i}_{M_i, 1}(\phi, \theta, 0) + \frac{1}{2 \sqrt{2}} \sin \theta \left[D^{J_i}_{M_i, -1}(\phi, \theta, 0) + D^{J_i}_{M_i, 1}(\phi, \theta, 0)\right]\Bigr\vert^2
\\ \nonumber
=& \frac{1}{8} \Bigr\vert D^{J_i}_{M_i, 1}(\phi, \theta, 0) - D^{J_i}_{M_i, -1}(\phi, \theta, 0)\Bigr\vert^2 \sin^2 \theta.
\end{align}
\end{widetext}
The second step of Eq. (\ref{eq:SigmaDerivation1}) uses Eq. (\ref{eq:DMatrix}) as well as the
summation (Clebsch-Gordan series)
\begin{widetext}
\begin{align}
\label{eq:Sum1}
& \sum_{J_k } (-1)^{M_k + \Omega_k}(2 J_k+1)  D^{J_k}_{M_k \Omega_k} (\phi,\theta,0)
 \begin{pmatrix}
  J_k & 1 & J_i\\
  -M_k &P& M_i
 \end{pmatrix}
 \begin{pmatrix}
  J_k & 1 & J_i\\
  -\Omega_k &Q& \Omega_i
 \end{pmatrix} =
 \\ \nonumber
 &  = \sum_{J_k}  D^{J_k \star}_{-M_k -\Omega_k}(\phi, \theta, 0)
  \begin{pmatrix}
  J_k & 1 & J_i\\
  -M_k &P& M_i
 \end{pmatrix}
 \begin{pmatrix}
  J_k & 1 & J_i\\
  -\Omega_k &Q& \Omega_i
 \end{pmatrix} =
 \\ \nonumber
 & = D^{J_i}_{M_i \Omega_i} D^{1}_{PQ}.
\end{align}
\end{widetext}

\subsection{Light polarization normal to the quantization axis}

With the assumptions $\{\Omega_i = 1,\Omega_k=0,M_k = M_i\pm1\}$, as for Sr$_2$ photodissociation from the $1_u$ states to the ground state continuum with light polarization normal to the quantization axis ($P=\pm1$), the axial-recoil cross section (\ref{eq:SigmaAR}) becomes
\begin{widetext}
\begin{align}
\label{eq:SigmaARAdapted2}
 \sigma(\theta, \phi)\propto \Bigr\vert\sum_{J_k P } (-1)^{M_k }(2 J_k+1)  D^{J_k}_{M_k 0} (\phi,\theta,0)
 \begin{pmatrix}
  J_k & 1 & J_i\\
  -M_k &P& M_i
 \end{pmatrix}\Bigr\vert^2.
\end{align}
\end{widetext}

\subsubsection{Partial waves restricted to $J_i$}

After restricting the partial waves to $J_k=J_i$, summing over $P$, and evaluating the $3j$ symbols, the cross section in Eq. (\ref{eq:SigmaARAdapted2}) becomes
\begin{widetext}
\begin{align}
\label{eq:SigmaDerivation2}
  \sigma(\theta, \phi) \propto & (2 J_i+1) \begin{pmatrix}
  J_i & 1 & J_i\\
  0 &-1& 1
 \end{pmatrix} \\ \nonumber \times & \bigr\vert D^{J_i}_{M_i+1, 0} (\phi,\theta,0) \sqrt{(J_i-M_i)(J_i+M_i+1)} - D^{J_i}_{M_i-1, 0} (\phi,\theta,0) \sqrt{(J_i+M_i)(J_i-M_i+1)} \bigr\vert^2.
\end{align}
\end{widetext}
We use the recursion formulas \cite{VarshalovichAngularMomentumBook}
\begin{widetext}
\begin{align}
\label{eq:Recursion1}
 D^{J_i}_{M_i+1, 0} (\phi,\theta,0) e^{i \phi} = \frac{\sqrt{J_i(J_i +1)}}{ \sqrt{(J_i-M_i)(J_i+M_i+1)}} \left[ \frac{1+\cos \theta}{2}  D^{J_i}_{M_i, -1}(\phi,\theta,0) -\frac{1-\cos \theta}{2}  D^{J_i}_{M_i, 1}(\phi,\theta,0) \right],
\end{align}
\begin{align}
\label{eq:Recursion2}
 D^{J_i}_{M_i-1, 0} (\phi,\theta,0)  e^{-i \phi} = \frac{\sqrt{J_i(J_i +1)}}{\sqrt{(J_i+M_i)(J_i-M_i+1)}} \left[ \frac{1+\cos \theta}{2}  D^{J_i}_{M_i, 1} (\phi,\theta,0) - \frac{1-\cos \theta}{2}  D^{J_i}_{M_i, -1}(\phi,\theta,0) \right]
\end{align}
\end{widetext}
to transform Eq. (\ref{eq:SigmaDerivation2}) into
\begin{widetext}
\begin{align}
\sigma(\theta, \phi) \propto & \bigr\vert D^{J_i}_{M_i, 1} (\phi,\theta,0) \times  (\cos \phi + i \cos \theta \sin \phi)
- D^{J_i}_{M_i, -1} (\phi,\theta,0) \times (\cos \phi - i \cos \theta \sin \phi)\bigr\vert^2
\\ \nonumber \propto \cos^2 \phi & \bigr\vert D^{J_i}_{M_i, 1} (\phi,\theta,0) - D^{J_i}_{M_i, -1} (\phi,\theta,0)\bigr\vert^2 + \cos^2 \theta \sin^2 \phi \bigr\vert D^{J_i}_{M_i, 1} (\phi,\theta,0) + D^{J_i}_{M_i, -1} (\phi,\theta,0)\bigr\vert^2.
 \end{align}
\end{widetext}

\subsubsection{Partial waves restricted to $J_i\pm1$}

With the partial waves limited to $J_k=J_i\pm1$, the cross section in Eq. (\ref{eq:SigmaARAdapted2}) can again be manipulated by taking the sum over all partial waves and subtracting the forbidden contributions,
\begin{widetext}
\begin{align}
\label{eq:SigmaDerivation3}
 \sigma(\theta, \phi)\propto & \Bigr\vert\sum_{J_k  P} (-1)^{M_k} (2 J_k+1)  D^{J_k}_{M_k 0} (\phi,\theta,0)
 \begin{pmatrix}
  J_k & 1 & J_i\\
  -M_k &P& M_i
 \end{pmatrix}
 \begin{pmatrix}
  J_k & 1 & J_i\\
  0 &-1& 1
 \end{pmatrix}
 \\ \nonumber -
 & (-1)^{M_k} (2 J_i+1)  D^{J_i}_{M_k 0} (\phi,\theta,0)
 \begin{pmatrix}
  J_i & 1 & J_i\\
  -M_k &P& M_i
 \end{pmatrix}
 \begin{pmatrix}
  J_i & 1 & J_i\\
  0 &-1& 1
  \end{pmatrix}
 \Bigr\vert^2.
\end{align}
\end{widetext}
After applying Eqs. (\ref{eq:DMatrix}) and (\ref{eq:Recursion1},\ref{eq:Recursion2}) to the first and second term on the right-hand side of Eq. (\ref{eq:SigmaDerivation3}), respectively, we obtain the cross section
\begin{widetext}
\begin{align}
\label{eq:SigmaDerivation4}
\nonumber
 \sigma(\theta, \phi)\propto& \bigr\vert D^{J_i}_{M_i, 1}(\phi,\theta,0) \left[D^{1}_{1, -1}(\phi,\theta,0) + D^{1}_{-1, -1} (\phi,\theta,0)\right] + \frac{1}{2}D^{J_i}_{M_i, 1} (\phi,\theta,0) \times  (\cos \phi + i \cos \theta \sin \phi)
 \\ \nonumber &
- \frac{1}{2} D^{J_i}_{M_i, -1} (\phi,\theta,0) \times (\cos \phi - i \cos \theta \sin \phi) \bigr\vert^2 =
 \\   = &
 \bigr\vert D^{J_i}_{M_i, 1} (\phi,\theta,0) \times  (\cos \phi + i \cos \theta \sin \phi)
+  D^{J_i}_{M_i, -1} (\phi,\theta,0) \times (\cos \phi - i \cos \theta \sin \phi) \bigr\vert^2
\\ \nonumber \propto & \cos^2 \phi \bigr\vert D^{J_i}_{M_i, 1} (\phi,\theta,0) + D^{J_i}_{M_i, -1} (\phi,\theta,0)\bigr\vert^2 + \cos^2 \theta \sin^2 \phi \bigr\vert D^{J_i}_{M_i, 1} (\phi,\theta,0) - D^{J_i}_{M_i, -1} (\phi,\theta,0)\bigr\vert^2.
\end{align}
\end{widetext}


\begin{thebibliography}{26}%
\makeatletter
\providecommand \@ifxundefined [1]{%
 \@ifx{#1\undefined}
}%
\providecommand \@ifnum [1]{%
 \ifnum #1\expandafter \@firstoftwo
 \else \expandafter \@secondoftwo
 \fi
}%
\providecommand \@ifx [1]{%
 \ifx #1\expandafter \@firstoftwo
 \else \expandafter \@secondoftwo
 \fi
}%
\providecommand \natexlab [1]{#1}%
\providecommand \enquote  [1]{``#1''}%
\providecommand \bibnamefont  [1]{#1}%
\providecommand \bibfnamefont [1]{#1}%
\providecommand \citenamefont [1]{#1}%
\providecommand \href@noop [0]{\@secondoftwo}%
\providecommand \href [0]{\begingroup \@sanitize@url \@href}%
\providecommand \@href[1]{\@@startlink{#1}\@@href}%
\providecommand \@@href[1]{\endgroup#1\@@endlink}%
\providecommand \@sanitize@url [0]{\catcode `\\12\catcode `\$12\catcode
  `\&12\catcode `\#12\catcode `\^12\catcode `\_12\catcode `\%12\relax}%
\providecommand \@@startlink[1]{}%
\providecommand \@@endlink[0]{}%
\providecommand \url  [0]{\begingroup\@sanitize@url \@url }%
\providecommand \@url [1]{\endgroup\@href {#1}{\urlprefix }}%
\providecommand \urlprefix  [0]{URL }%
\providecommand \Eprint [0]{\href }%
\providecommand \doibase [0]{http://dx.doi.org/}%
\providecommand \selectlanguage [0]{\@gobble}%
\providecommand \bibinfo  [0]{\@secondoftwo}%
\providecommand \bibfield  [0]{\@secondoftwo}%
\providecommand \translation [1]{[#1]}%
\providecommand \BibitemOpen [0]{}%
\providecommand \bibitemStop [0]{}%
\providecommand \bibitemNoStop [0]{.\EOS\space}%
\providecommand \EOS [0]{\spacefactor3000\relax}%
\providecommand \BibitemShut  [1]{\csname bibitem#1\endcsname}%
\let\auto@bib@innerbib\@empty
\bibitem [{\citenamefont
  {Balakrishnan}(2016)}]{BalakrishnanJCP16_UltracoldMolecules}%
  \BibitemOpen
  \bibfield  {author} {\bibinfo {author} {\bibfnamefont {N.}~\bibnamefont
  {Balakrishnan}},\ }\bibfield  {title} {\enquote {\bibinfo {title}
  {{Perspective: Ultracold molecules and the dawn of cold controlled
  chemistry}},}\ }\href@noop {} {\bibfield  {journal} {\bibinfo  {journal} {J.
  Chem. Phys.}\ }\textbf {\bibinfo {volume} {145}},\ \bibinfo {pages} {150901}
  (\bibinfo {year} {2016})}\BibitemShut {NoStop}%
\bibitem [{\citenamefont {Sato}(2001)}]{SatoCR01_PhotodissociationReview}%
  \BibitemOpen
  \bibfield  {author} {\bibinfo {author} {\bibfnamefont {H.}~\bibnamefont
  {Sato}},\ }\bibfield  {title} {\enquote {\bibinfo {title} {Photodissociation
  of simple molecules in the gas phase},}\ }\href@noop {} {\bibfield  {journal}
  {\bibinfo  {journal} {Chem. Rev.}\ }\textbf {\bibinfo {volume} {101}},\
  \bibinfo {pages} {2687--2725} (\bibinfo {year} {2001})}\BibitemShut {NoStop}%
\bibitem [{\citenamefont {Zare}\ and\ \citenamefont
  {Herschbach}(1963)}]{HerschbachZarePIEEE63_DiatomicPhotodissociation}%
  \BibitemOpen
  \bibfield  {author} {\bibinfo {author} {\bibfnamefont {R.~N.}\ \bibnamefont
  {Zare}}\ and\ \bibinfo {author} {\bibfnamefont {D.~R.}\ \bibnamefont
  {Herschbach}},\ }\bibfield  {title} {\enquote {\bibinfo {title} {Doppler line
  shape of atomic fluorescence excited by molecular photodissociation},}\
  }\href@noop {} {\bibfield  {journal} {\bibinfo  {journal} {Proc. IEEE}\
  }\textbf {\bibinfo {volume} {51}},\ \bibinfo {pages} {173--182} (\bibinfo
  {year} {1963})}\BibitemShut {NoStop}%
\bibitem [{\citenamefont {Zare}(1972)}]{ZareMPC72_PhotoejectionDynamics}%
  \BibitemOpen
  \bibfield  {author} {\bibinfo {author} {\bibfnamefont {R.~N.}\ \bibnamefont
  {Zare}},\ }\bibfield  {title} {\enquote {\bibinfo {title} {Photoejection
  dynamics},}\ }\href@noop {} {\bibfield  {journal} {\bibinfo  {journal} {Mol.
  Photochem.}\ }\textbf {\bibinfo {volume} {4}},\ \bibinfo {pages} {1--37}
  (\bibinfo {year} {1972})}\BibitemShut {NoStop}%
\bibitem [{\citenamefont {Choi}\ and\ \citenamefont
  {Bernstein}(1986)}]{BernsteinChoiJCP86_StateSelectedPhotofragmentation}%
  \BibitemOpen
  \bibfield  {author} {\bibinfo {author} {\bibfnamefont {S.~E.}\ \bibnamefont
  {Choi}}\ and\ \bibinfo {author} {\bibfnamefont {R.~B.}\ \bibnamefont
  {Bernstein}},\ }\bibfield  {title} {\enquote {\bibinfo {title} {Theory of
  oriented symmetric-top molecule beams: Precession, degree of orientation, and
  photofragmentation of rotationally state-selected molecules},}\ }\href@noop
  {} {\bibfield  {journal} {\bibinfo  {journal} {J. Chem. Phys.}\ }\textbf
  {\bibinfo {volume} {85}},\ \bibinfo {pages} {150--161} (\bibinfo {year}
  {1986})}\BibitemShut {NoStop}%
\bibitem [{\citenamefont
  {Zare}(1989)}]{ZareCPL89_PhotofragmentAngularDistributions}%
  \BibitemOpen
  \bibfield  {author} {\bibinfo {author} {\bibfnamefont {R.~N.}\ \bibnamefont
  {Zare}},\ }\bibfield  {title} {\enquote {\bibinfo {title} {Photofragment
  angular distributions from oriented symmetric-top precursor molecules},}\
  }\href@noop {} {\bibfield  {journal} {\bibinfo  {journal} {Chem. Phys.
  Lett.}\ }\textbf {\bibinfo {volume} {156}},\ \bibinfo {pages} {1--6}
  (\bibinfo {year} {1989})}\BibitemShut {NoStop}%
\bibitem [{\citenamefont
  {Seideman}(1996)}]{SeidemanCPL96_MagneticStateSelectedPDDistributions}%
  \BibitemOpen
  \bibfield  {author} {\bibinfo {author} {\bibfnamefont {T.}~\bibnamefont
  {Seideman}},\ }\bibfield  {title} {\enquote {\bibinfo {title} {The analysis
  of magnetic-state-selected angular distributions: a quantum mechanical form
  and an asymptotic approximation},}\ }\href@noop {} {\bibfield  {journal}
  {\bibinfo  {journal} {Chem. Phys. Lett.}\ }\textbf {\bibinfo {volume}
  {253}},\ \bibinfo {pages} {279--285} (\bibinfo {year} {1996})}\BibitemShut
  {NoStop}%
\bibitem [{\citenamefont {Beswick}\ and\ \citenamefont
  {Zare}(2008)}]{ZareBeswickJCP08_PhotofragmentAngularDistrQuantClass}%
  \BibitemOpen
  \bibfield  {author} {\bibinfo {author} {\bibfnamefont {J.~A.}\ \bibnamefont
  {Beswick}}\ and\ \bibinfo {author} {\bibfnamefont {R.~N.}\ \bibnamefont
  {Zare}},\ }\bibfield  {title} {\enquote {\bibinfo {title} {On the quantum and
  quasiclassical angular distributions of photofragments},}\ }\href@noop {}
  {\bibfield  {journal} {\bibinfo  {journal} {J. Chem. Phys.}\ }\textbf
  {\bibinfo {volume} {129}},\ \bibinfo {pages} {164315} (\bibinfo {year}
  {2008})}\BibitemShut {NoStop}%
\bibitem [{\citenamefont {McDonald}\ \emph {et~al.}(2016)\citenamefont
  {McDonald}, \citenamefont {McGuyer}, \citenamefont {Apfelbeck}, \citenamefont
  {Lee}, \citenamefont {Majewska}, \citenamefont {Moszynski},\ and\
  \citenamefont {Zelevinsky}}]{ZelevinskyMcDonaldNature16_Sr2PD}%
  \BibitemOpen
  \bibfield  {author} {\bibinfo {author} {\bibfnamefont {M.}~\bibnamefont
  {McDonald}}, \bibinfo {author} {\bibfnamefont {B.~H.}\ \bibnamefont
  {McGuyer}}, \bibinfo {author} {\bibfnamefont {F.}~\bibnamefont {Apfelbeck}},
  \bibinfo {author} {\bibfnamefont {C.-H.}\ \bibnamefont {Lee}}, \bibinfo
  {author} {\bibfnamefont {I.}~\bibnamefont {Majewska}}, \bibinfo {author}
  {\bibfnamefont {R.}~\bibnamefont {Moszynski}}, \ and\ \bibinfo {author}
  {\bibfnamefont {T.}~\bibnamefont {Zelevinsky}},\ }\bibfield  {title}
  {\enquote {\bibinfo {title} {{Photodissociation of ultracold diatomic
  strontium molecules with quantum state control}},}\ }\href@noop {} {\bibfield
   {journal} {\bibinfo  {journal} {Nature}\ }\textbf {\bibinfo {volume}
  {534}},\ \bibinfo {pages} {122--126} (\bibinfo {year} {2016})}\BibitemShut
  {NoStop}%
\bibitem [{\citenamefont {McDonald}\ \emph {et~al.}(2018)\citenamefont
  {McDonald}, \citenamefont {Majewska}, \citenamefont {Lee}, \citenamefont
  {Kondov}, \citenamefont {McGuyer}, \citenamefont {Moszynski},\ and\
  \citenamefont {Zelevinsky}}]{ZelevinskyMcDonaldPRL18_PDMagneticField}%
  \BibitemOpen
  \bibfield  {author} {\bibinfo {author} {\bibfnamefont {M.}~\bibnamefont
  {McDonald}}, \bibinfo {author} {\bibfnamefont {I.}~\bibnamefont {Majewska}},
  \bibinfo {author} {\bibfnamefont {C.-H.}\ \bibnamefont {Lee}}, \bibinfo
  {author} {\bibfnamefont {S.~S.}\ \bibnamefont {Kondov}}, \bibinfo {author}
  {\bibfnamefont {B.~H.}\ \bibnamefont {McGuyer}}, \bibinfo {author}
  {\bibfnamefont {R.}~\bibnamefont {Moszynski}}, \ and\ \bibinfo {author}
  {\bibfnamefont {T.}~\bibnamefont {Zelevinsky}},\ }\bibfield  {title}
  {\enquote {\bibinfo {title} {Control of ultracold photodissociation with
  magnetic fields},}\ }\href@noop {} {\bibfield  {journal} {\bibinfo  {journal}
  {Phys. Rev. Lett.}\ }\textbf {\bibinfo {volume} {120}},\ \bibinfo {pages}
  {033201} (\bibinfo {year} {2018})}\BibitemShut {NoStop}%
\bibitem [{\citenamefont {Skomorowski}\ \emph {et~al.}(2012)\citenamefont
  {Skomorowski}, \citenamefont {Paw{\l}owski}, \citenamefont {Koch},\ and\
  \citenamefont {Moszynski}}]{MoszynskiSkomorowskiJCP12_Sr2Dynamics}%
  \BibitemOpen
  \bibfield  {author} {\bibinfo {author} {\bibfnamefont {W.}~\bibnamefont
  {Skomorowski}}, \bibinfo {author} {\bibfnamefont {F.}~\bibnamefont
  {Paw{\l}owski}}, \bibinfo {author} {\bibfnamefont {C.~P.}\ \bibnamefont
  {Koch}}, \ and\ \bibinfo {author} {\bibfnamefont {R.}~\bibnamefont
  {Moszynski}},\ }\bibfield  {title} {\enquote {\bibinfo {title}
  {{Rovibrational dynamics of the strontium molecule in the A$^1\Sigma_u^+$,
  c$^3\Pi_u$, and a$^3\Sigma_u^+$ manifold from state-of-the-art \textit{ab
  initio} calculations}},}\ }\href@noop {} {\bibfield  {journal} {\bibinfo
  {journal} {J. Chem. Phys.}\ }\textbf {\bibinfo {volume} {136}},\ \bibinfo
  {pages} {194306} (\bibinfo {year} {2012})}\BibitemShut {NoStop}%
\bibitem [{\citenamefont {Borkowski}\ \emph {et~al.}(2014)\citenamefont
  {Borkowski}, \citenamefont {Morzy\'{n}ski}, \citenamefont {Ciury{\l}o},
  \citenamefont {Julienne}, \citenamefont {Yan}, \citenamefont {DeSalvo},\ and\
  \citenamefont {Killian}}]{KillianBorkowskiPRA14_SrPAMassScaling}%
  \BibitemOpen
  \bibfield  {author} {\bibinfo {author} {\bibfnamefont {M.}~\bibnamefont
  {Borkowski}}, \bibinfo {author} {\bibfnamefont {P.}~\bibnamefont
  {Morzy\'{n}ski}}, \bibinfo {author} {\bibfnamefont {R.}~\bibnamefont
  {Ciury{\l}o}}, \bibinfo {author} {\bibfnamefont {P.~S.}\ \bibnamefont
  {Julienne}}, \bibinfo {author} {\bibfnamefont {M.}~\bibnamefont {Yan}},
  \bibinfo {author} {\bibfnamefont {B.~J.}\ \bibnamefont {DeSalvo}}, \ and\
  \bibinfo {author} {\bibfnamefont {T.~C.}\ \bibnamefont {Killian}},\
  }\bibfield  {title} {\enquote {\bibinfo {title} {{Mass scaling and
  nonadiabatic effects in photoassociation spectroscopy of ultracold strontium
  atoms}},}\ }\href@noop {} {\bibfield  {journal} {\bibinfo  {journal} {Phys.
  Rev. A}\ }\textbf {\bibinfo {volume} {90}},\ \bibinfo {pages} {032713}
  (\bibinfo {year} {2014})}\BibitemShut {NoStop}%
\bibitem [{\citenamefont {McGuyer}\ \emph {et~al.}(2013)\citenamefont
  {McGuyer}, \citenamefont {Osborn}, \citenamefont {McDonald}, \citenamefont
  {Reinaudi}, \citenamefont {Skomorowski}, \citenamefont {Moszynski},\ and\
  \citenamefont {Zelevinsky}}]{ZelevinskyMcGuyerPRL13_Sr2ZeemanNonadiabatic}%
  \BibitemOpen
  \bibfield  {author} {\bibinfo {author} {\bibfnamefont {B.~H.}\ \bibnamefont
  {McGuyer}}, \bibinfo {author} {\bibfnamefont {C.~B.}\ \bibnamefont {Osborn}},
  \bibinfo {author} {\bibfnamefont {M.}~\bibnamefont {McDonald}}, \bibinfo
  {author} {\bibfnamefont {G.}~\bibnamefont {Reinaudi}}, \bibinfo {author}
  {\bibfnamefont {W.}~\bibnamefont {Skomorowski}}, \bibinfo {author}
  {\bibfnamefont {R.}~\bibnamefont {Moszynski}}, \ and\ \bibinfo {author}
  {\bibfnamefont {T.}~\bibnamefont {Zelevinsky}},\ }\bibfield  {title}
  {\enquote {\bibinfo {title} {Nonadiabatic effects in ultracold molecules via
  anomalous linear and quadratic {Zeeman} shifts},}\ }\href@noop {} {\bibfield
  {journal} {\bibinfo  {journal} {Phys. Rev. Lett.}\ }\textbf {\bibinfo
  {volume} {111}},\ \bibinfo {pages} {243003} (\bibinfo {year}
  {2013})}\BibitemShut {NoStop}%
\bibitem [{\citenamefont {McGuyer}\ \emph
  {et~al.}(2015{\natexlab{a}})\citenamefont {McGuyer}, \citenamefont
  {McDonald}, \citenamefont {Iwata}, \citenamefont {Skomorowski}, \citenamefont
  {Moszynski},\ and\ \citenamefont
  {Zelevinsky}}]{ZelevinskyMcGuyerPRL15_Sr2ForbiddenE1}%
  \BibitemOpen
  \bibfield  {author} {\bibinfo {author} {\bibfnamefont {B.~H.}\ \bibnamefont
  {McGuyer}}, \bibinfo {author} {\bibfnamefont {M.}~\bibnamefont {McDonald}},
  \bibinfo {author} {\bibfnamefont {G.~Z.}\ \bibnamefont {Iwata}}, \bibinfo
  {author} {\bibfnamefont {W.}~\bibnamefont {Skomorowski}}, \bibinfo {author}
  {\bibfnamefont {R.}~\bibnamefont {Moszynski}}, \ and\ \bibinfo {author}
  {\bibfnamefont {T.}~\bibnamefont {Zelevinsky}},\ }\bibfield  {title}
  {\enquote {\bibinfo {title} {{Control of optical transitions with magnetic
  fields in weakly bound molecules}},}\ }\href@noop {} {\bibfield  {journal}
  {\bibinfo  {journal} {Phys. Rev. Lett.}\ }\textbf {\bibinfo {volume} {115}},\
  \bibinfo {pages} {053001} (\bibinfo {year} {2015}{\natexlab{a}})}\BibitemShut
  {NoStop}%
\bibitem [{\citenamefont {McGuyer}\ \emph
  {et~al.}(2015{\natexlab{b}})\citenamefont {McGuyer}, \citenamefont
  {McDonald}, \citenamefont {Iwata}, \citenamefont {Tarallo}, \citenamefont
  {Grier}, \citenamefont {Apfelbeck},\ and\ \citenamefont
  {Zelevinsky}}]{ZelevinskyMcGuyerNJP15_Sr2Spectroscopy}%
  \BibitemOpen
  \bibfield  {author} {\bibinfo {author} {\bibfnamefont {B.~H.}\ \bibnamefont
  {McGuyer}}, \bibinfo {author} {\bibfnamefont {M.}~\bibnamefont {McDonald}},
  \bibinfo {author} {\bibfnamefont {G.~Z.}\ \bibnamefont {Iwata}}, \bibinfo
  {author} {\bibfnamefont {M.~G.}\ \bibnamefont {Tarallo}}, \bibinfo {author}
  {\bibfnamefont {A.~T.}\ \bibnamefont {Grier}}, \bibinfo {author}
  {\bibfnamefont {F.}~\bibnamefont {Apfelbeck}}, \ and\ \bibinfo {author}
  {\bibfnamefont {T.}~\bibnamefont {Zelevinsky}},\ }\bibfield  {title}
  {\enquote {\bibinfo {title} {{High-precision spectroscopy of ultracold
  molecules in an optical lattice}},}\ }\href@noop {} {\bibfield  {journal}
  {\bibinfo  {journal} {New J. Phys.}\ }\textbf {\bibinfo {volume} {17}},\
  \bibinfo {pages} {055004} (\bibinfo {year} {2015}{\natexlab{b}})}\BibitemShut
  {NoStop}%
\bibitem [{\citenamefont {McGuyer}\ \emph
  {et~al.}(2015{\natexlab{c}})\citenamefont {McGuyer}, \citenamefont
  {McDonald}, \citenamefont {Iwata}, \citenamefont {Tarallo}, \citenamefont
  {Skomorowski}, \citenamefont {Moszynski},\ and\ \citenamefont
  {Zelevinsky}}]{ZelevinskyMcGuyerNPhys15_Sr2M1}%
  \BibitemOpen
  \bibfield  {author} {\bibinfo {author} {\bibfnamefont {B.~H.}\ \bibnamefont
  {McGuyer}}, \bibinfo {author} {\bibfnamefont {M.}~\bibnamefont {McDonald}},
  \bibinfo {author} {\bibfnamefont {G.~Z.}\ \bibnamefont {Iwata}}, \bibinfo
  {author} {\bibfnamefont {M.~G.}\ \bibnamefont {Tarallo}}, \bibinfo {author}
  {\bibfnamefont {W.}~\bibnamefont {Skomorowski}}, \bibinfo {author}
  {\bibfnamefont {R.}~\bibnamefont {Moszynski}}, \ and\ \bibinfo {author}
  {\bibfnamefont {T.}~\bibnamefont {Zelevinsky}},\ }\bibfield  {title}
  {\enquote {\bibinfo {title} {{Precise study of asymptotic physics with
  subradiant ultracold molecules}},}\ }\href@noop {} {\bibfield  {journal}
  {\bibinfo  {journal} {Nature Phys.}\ }\textbf {\bibinfo {volume} {11}},\
  \bibinfo {pages} {32--36} (\bibinfo {year} {2015}{\natexlab{c}})}\BibitemShut
  {NoStop}%
\bibitem [{\citenamefont {Kondov}\ \emph {et~al.}(2018)\citenamefont {Kondov},
  \citenamefont {Lee}, \citenamefont {McDonald}, \citenamefont {McGuyer},
  \citenamefont {Majewska}, \citenamefont {Moszynski},\ and\ \citenamefont
  {Zelevinsky}}]{ZelevinskyKondovArxiv18_QMQCPD}%
  \BibitemOpen
  \bibfield  {author} {\bibinfo {author} {\bibfnamefont {S.~S.}\ \bibnamefont
  {Kondov}}, \bibinfo {author} {\bibfnamefont {C.-H.}\ \bibnamefont {Lee}},
  \bibinfo {author} {\bibfnamefont {M.}~\bibnamefont {McDonald}}, \bibinfo
  {author} {\bibfnamefont {B.~H.}\ \bibnamefont {McGuyer}}, \bibinfo {author}
  {\bibfnamefont {I.}~\bibnamefont {Majewska}}, \bibinfo {author}
  {\bibfnamefont {R.}~\bibnamefont {Moszynski}}, \ and\ \bibinfo {author}
  {\bibfnamefont {T.}~\bibnamefont {Zelevinsky}},\ }\bibfield  {title}
  {\enquote {\bibinfo {title} {Crossover from the ultracold to the
  quasiclassical regime in state-selected photodissociation},}\ }\href@noop {}
  {\bibfield  {journal} {\bibinfo  {journal} {arXiv:1805.08850}\ } (\bibinfo
  {year} {2018})}\BibitemShut {NoStop}%
\bibitem [{\citenamefont {Reinaudi}\ \emph {et~al.}(2012)\citenamefont
  {Reinaudi}, \citenamefont {Osborn}, \citenamefont {McDonald}, \citenamefont
  {Kotochigova},\ and\ \citenamefont
  {Zelevinsky}}]{ZelevinskyReinaudiPRL12_Sr2}%
  \BibitemOpen
  \bibfield  {author} {\bibinfo {author} {\bibfnamefont {G.}~\bibnamefont
  {Reinaudi}}, \bibinfo {author} {\bibfnamefont {C.~B.}\ \bibnamefont
  {Osborn}}, \bibinfo {author} {\bibfnamefont {M.}~\bibnamefont {McDonald}},
  \bibinfo {author} {\bibfnamefont {S.}~\bibnamefont {Kotochigova}}, \ and\
  \bibinfo {author} {\bibfnamefont {T.}~\bibnamefont {Zelevinsky}},\ }\bibfield
   {title} {\enquote {\bibinfo {title} {{Optical production of stable ultracold
  $^{88}$Sr$_2$ molecules}},}\ }\href@noop {} {\bibfield  {journal} {\bibinfo
  {journal} {Phys. Rev. Lett.}\ }\textbf {\bibinfo {volume} {109}},\ \bibinfo
  {pages} {115303} (\bibinfo {year} {2012})}\BibitemShut {NoStop}%
\bibitem [{\citenamefont {Child}(2010)}]{ChildBook}%
  \BibitemOpen
  \bibfield  {author} {\bibinfo {author} {\bibfnamefont {M.~S.}\ \bibnamefont
  {Child}},\ }\href@noop {} {\emph {\bibinfo {title} {Molecular Collision
  Theory}}}\ (\bibinfo  {publisher} {Dover Publications},\ \bibinfo {address}
  {Mineola, New York},\ \bibinfo {year} {2010})\BibitemShut {NoStop}%
\bibitem [{\citenamefont {Nikitin}\ and\ \citenamefont
  {Umanskii}(1984)}]{NikitinUmanskiiBook}%
  \BibitemOpen
  \bibfield  {author} {\bibinfo {author} {\bibfnamefont {E.~E.}\ \bibnamefont
  {Nikitin}}\ and\ \bibinfo {author} {\bibfnamefont {S.~Y.}\ \bibnamefont
  {Umanskii}},\ }\href@noop {} {\emph {\bibinfo {title} {Theory of Slow Atomic
  Collisions}}}\ (\bibinfo  {publisher} {Springer-Verlag},\ \bibinfo {address}
  {Berlin Heidelberg},\ \bibinfo {year} {1984})\BibitemShut {NoStop}%
\bibitem [{\citenamefont {Wrede}\ \emph {et~al.}(2002)\citenamefont {Wrede},
  \citenamefont {Wouters}, \citenamefont {Beckert}, \citenamefont {Dixon},\
  and\ \citenamefont {Ashfold}}]{AshfoldWredeJCP02_AxialRecoilBreakdown}%
  \BibitemOpen
  \bibfield  {author} {\bibinfo {author} {\bibfnamefont {E.}~\bibnamefont
  {Wrede}}, \bibinfo {author} {\bibfnamefont {E.~R.}\ \bibnamefont {Wouters}},
  \bibinfo {author} {\bibfnamefont {M.}~\bibnamefont {Beckert}}, \bibinfo
  {author} {\bibfnamefont {R.~N.}\ \bibnamefont {Dixon}}, \ and\ \bibinfo
  {author} {\bibfnamefont {M.~N.~R.}\ \bibnamefont {Ashfold}},\ }\bibfield
  {title} {\enquote {\bibinfo {title} {{Quasiclassical and quantum mechanical
  modeling of the breakdown of the axial recoil approximation observed in the
  near threshold photolysis of IBr and Br$_2$}},}\ }\href@noop {} {\bibfield
  {journal} {\bibinfo  {journal} {J. Chem. Phys.}\ }\textbf {\bibinfo {volume}
  {116}},\ \bibinfo {pages} {6064} (\bibinfo {year} {2002})}\BibitemShut
  {NoStop}%
\bibitem [{Note1()}]{Note1}%
  \BibitemOpen
  \bibinfo {note} {In Ref. \cite {AshfoldWredeJCP02_AxialRecoilBreakdown} the
  semiclassical model is referred to as ``quasiclassical''. However, it is not
  equivalent to the quasiclassical description from Refs. \cite
  {BernsteinChoiJCP86_StateSelectedPhotofragmentation,ZareCPL89_PhotofragmentAngularDistributions},
  and is a more accurate approximation.}\BibitemShut {Stop}%
\bibitem [{Note2()}]{Note2}%
  \BibitemOpen
  \bibinfo {note} {For the WKB images in Fig. \ref {fig:PADs}, the axial-recoil
  value of $R$ was used, since $R$ approaches the high-energy limit faster than
  $\delta $.}\BibitemShut {Stop}%
\bibitem [{Note3()}]{Note3}%
  \BibitemOpen
  \bibinfo {note} {For $M_i=P=0$, quantum statistics (when combined with the
  properties of the $3j$ symbol in Eqs. (\ref {eq:SigmaQM},\ref {eq:SigmaAR}))
  can cause a normally allowed photodissociation pathway to become forbidden,
  for example for initial states $1_u(v,J_i,0)$ with even $J_i$ in bosonic-Sr
  dimers. Such transitions, however, can be enabled with weak magnetic fields
  \cite
  {ZelevinskyMcDonaldNature16_Sr2PD,ZelevinskyMcDonaldPRL18_PDMagneticField}.}\BibitemShut
  {Stop}%
\bibitem [{Note4()}]{Note4}%
  \BibitemOpen
  \bibinfo {note} {Two other configurations with this property,
  $1_u(J_i=1,M_i=1,P=0)$ and $1_u(J_i=3,M_i=3,P=0)$, are not
  shown.}\BibitemShut {Stop}%
\bibitem [{\citenamefont {Varshalovich}\ \emph {et~al.}(1988)\citenamefont
  {Varshalovich}, \citenamefont {Moskalev},\ and\ \citenamefont
  {Khersonskii}}]{VarshalovichAngularMomentumBook}%
  \BibitemOpen
  \bibfield  {author} {\bibinfo {author} {\bibfnamefont {D.~A.}\ \bibnamefont
  {Varshalovich}}, \bibinfo {author} {\bibfnamefont {A.~N.}\ \bibnamefont
  {Moskalev}}, \ and\ \bibinfo {author} {\bibfnamefont {V.~K.}\ \bibnamefont
  {Khersonskii}},\ }\href@noop {} {\emph {\bibinfo {title} {Quantum Theory of
  Angular Momentum}}}\ (\bibinfo  {publisher} {World Scientific},\ \bibinfo
  {address} {Singapore},\ \bibinfo {year} {1988})\BibitemShut {NoStop}%
\end{thebibliography}

\end{document}